\documentclass[a4paper,fleqn,usenatbib]{mnras}

\usepackage{newtxtext,newtxmath}

\usepackage[T1]{fontenc}
\usepackage{ae,aecompl}
\usepackage{footnote}
\usepackage{caption}
\usepackage{subcaption}
\usepackage{graphicx}
\usepackage{wrapfig}
\usepackage{lscape}
\usepackage{rotating}
\usepackage{epstopdf}

\usepackage{graphicx}	
\usepackage{amsmath,bm}	
\usepackage{lmodern}

\usepackage{tikz,xcolor,hyperref}

\definecolor{lime}{HTML}{A6CE39}
\DeclareRobustCommand{\orcidicon}{
	\begin{tikzpicture}
	\draw[lime, fill=lime] (0,0) 
	circle [radius=0.16] 
	node[white] {{\fontfamily{qag}\selectfont \tiny ID}};
	\draw[white, fill=white] (-0.0625,0.095) 
	circle [radius=0.007];
	\end{tikzpicture}
	\hspace{-2mm}
}

\foreach \x in {A, ..., Z}{\expandafter\xdef\csname orcid\x\endcsname{\noexpand\href{https://orcid.org/\csname orcidauthor\x\endcsname}
			{\noexpand\orcidicon}}
}

\def\mnras{MNRAS}
\def\apj{ApJ}
\def\apjs{ApJS}
\def\aap{A\&A}
\def\aaps{A\&AS}
\def\aj{AJ}
\def\pasp{PASP}

\usepackage[normalem]{ulem}  
\newcommand\redout{\bgroup\markoverwith
{\textcolor{red}{\rule[0.5ex]{2pt}{0.8pt}}}\ULon}
\newcommand{\ha}{H$\alpha$}
\newcommand{\hb}{H$\beta$}

\newcommand{\oi}{[O\,{\sc i}]}

\newcommand{\oiii}{[O\,{\sc iii}]}

\newcommand{\nii}{[N\,{\sc ii}]}

\newcommand{\sii}{[S\,{\sc ii}]}
\def \st{\ifmmode{^{\mathrm{st}}}\else{${^{\mathrm{st}}}$}\fi}
\def \nd{\ifmmode{^{\mathrm{nd}}}\else{${^{\mathrm{nd}}}$}\fi}
\def \rd{\ifmmode{^{\mathrm{rd}}}\else{${^{\mathrm{rd}}}$}\fi}
\def \th{\ifmmode{^{\mathrm{th}}}\else{${^{\mathrm{th}}}$}\fi}
\newcommand{\hnii}{{\rm H}$\alpha+$[N~{\sc ii}]}
\newcommand{\h}{$^{\rm h}$}
\newcommand{\m}{$^{\rm m}$}
\newcommand{\s}{$^{\rm s}$}

\newcommand{\hbeta}{\rm H$\beta$}

\newcommand{\flux}{$10^{-17}$ erg s$^{-1}$ cm$^{-2}$ arcsec$^{-2}$}
\newcommand{\fluxa}{$10^{-17}$ erg s$^{-1}$ cm$^{-2}$ arcsec$^{-2}$}

\newcommand{\vel}{\rm km s$^{-1}$}
\newcommand{\sulfur}{[S~{\sc ii}]}
\newcommand{\nitrogen}{[N~{\sc ii}]}
\newcommand{\oxygen}{[O~{\sc iii}]}
\newcommand{\siirat}{[\ion{S}{ii}]~$\lambda\lambda$~6716/6731} 
\newcommand{\HNII}{{\rm H}$\alpha+$[N {\sc ii}]~6548, 6584~\AA}  
\newcommand{\OIII}{[O {\sc iii}]~5007~\AA}
\newcommand{\SII}{[S~{\sc ii}]~6716, 6731~\AA}
\newcommand{\Ha}{H$\alpha$}

\newcommand{\NI}{[N{\sc i}]}

\newcommand{\OI}{[O{\sc i}]}

\title[Deep optical study of the MM SNR G~132.7$+$1.3 (HB3)]
{Deep optical study of the mixed-morphology supernova remnant G~132.7$+$1.3 (HB3)}

\author[P. Boumis et al.]{P. Boumis,$^{1\orcidA{}}$\thanks{ptb@astro.noa.gr}
A. Chiotellis,$^{1\orcidB{}}$ V. Fragkou,$^{2}$ S.Akras,$^{1\orcidC{}}$, S. Derlopa,$^{1}$ M. Kopsacheili,$^{3,4\orcidD{}}$ 
\newauthor I. Leonidaki,$^{3,4\orcidD{}}$ 
J. Alikakos,$^{1}$ 
E. V. Palaiologou$^{3,4}$
E. Harvey,$^{5\orcidE{}}$ 
D. Souropanis$^{1,6\orcidF{}}$ 
\\
$^{1}$Institute for Astronomy, Astrophysics, Space Applications
and Remote Sensing, National Observatory of Athens,
15236 Penteli, Greece\\
$^{2}$Instituto de Astronom\'ia, Universidad Nacional Aut\'onoma de M\'exico, Ensenada 22800, Baja California, Mexico\\ 
$^{3}$Institute of Astrophysics, Foundation for Research and Technology-Hellas, 71110 Heraklion, Crete, Greece\\
$^{4}$Department of Physics, University of Crete, GR-71003, Heraklion, Crete, Greece\\
$^{5}$Astrophysics Research Institute, Liverpool John Moores University, IC2, Liverpool Science Park, Liverpool, L3 5RF, UK\\
$^{6}$Department of Physics, National and Kapodistrian University of Athens, Panepistimiopolis, 15784 Zografos, Greece\\
}

\date{Accepted 2022 February 08. Received 2022 February 02; in original form 2021 July 08}

\pubyear{2022}

\begin{document}
\label{firstpage}
\pagerange{\pageref{firstpage}--\pageref{lastpage}}
\maketitle

\begin{abstract}
We present optical ccd images of the large supernova remnant (SNR) G132.7$+$1.3 (HB3) covering its full extent for the first time, in the emission lines of \hnii, \sulfur\ and \oiii, where new and known filamentary and diffuse structures are detected. These observations are supplemented by new low-resolution long-slit spectra and higher-resolution images in the same emission lines. Both the flux-calibrated images and spectra confirm that the optical emission originates from shock-heated gas since the \sulfur/\Ha\ $>$ 0.4. Our findings are also consistent with the recently developed emission line ratio diagnostics for distinguishing SNRs from {H\sc ii} regions. A multi-wavelength comparison among our optical data and relevant observations in radio, X-rays, $\gamma$-rays and CO bands, provided additional evidence on the interaction of HB3 with the surrounding clouds and clarified the borders of the SNR and the adjacent cloud. We discuss the supernova (SN) properties and evolution that led to the current observables of HB3 and we show that the remnant has most likely passed at the pressure~driven~snowplow~phase. The estimated SN energy was found to be $\left(3.7 \pm 1.5\right) \times 10^{51}$~erg and the current SNR age $\left(5.1 \pm 2.1\right) \times 10^4$~yrs. We present an alternative scenario according to which the SNR evolved in the wind bubble cavity excavated by the progenitor star and currently is interacting with its density walls. We show that the overall mixed morphology properties of HB3 can be explained if the SN resulted by a Wolf-Rayet progenitor star with mass $\sim 34 \rm~M_{\rm\odot}$.  
\end{abstract}



\section{Introduction}

Supernova remnants (SNRs) are the aftermath of supernova (SN) explosions, that result from the interaction of the supersonically moving stellar ejecta with the ambient medium. The resultant nebula is an excellent celestial lab from which we can infer the mechanisms responsible for the explosive death of certain stars, the shock wave physics and the local ambient medium properties. In addition, SNRs are considered one of the most efficient cosmic-ray accelerators while they substantially contribute on the chemical and dynamical evolution of their host galaxies. For these reasons several targeted, multiwavelength observational campaigns have been conducted aiming to decipher the formation and evolution processes of these nebulae. While most SNRs have been detected in radio wavelengths (see \citealt{Dubner2015}; \citealt{Green2017}), optical observation of SNRs allow us to inspect the SNR properties such  as the filamentary morphology, the chemical abundances and thermodynamical properties of the shocked gas, the remnant’s expansion velocities and the density of the ambient medium (e.g   \citealt{Boumis2008,Boumis2009}; \citealt{Alikakos2012,Stupar2012,Sabin2013};  \citealt{Stupar2018,How2018}; \citealt{Fesen2020,Fesen2021}).

The SNRs in our Galaxy can be very extended structures depending on their distance, their evolutionary stage,  the characteristics of their parent stars and the properties of the ISM that encloses them (see e.g. \citealt{Asvarov2014}). G132.7$+$1.3 (also known as HB3 radio source) is among the largest Galactic SNRs observed so far \footnote{Only recently optical SNRs were found with a few degrees sizes at high Galactic latitudes; \citep{Fesen2020,Fesen2021}} with an angular size of $90\arcmin \times 123\arcmin$ in 408 MHz radio maps \citep{Caswell1967} which implies a physical size of 52 $\times$ 72 $d_2$ pc, where $d_2$ is the distance of the remnant in units of 2 kpc. The distance of HB3 has been suggested to be within the range of   $\sim 1.6 - 2.4$~kpc \citep{Routledge1991, Fesen1995, Xu2006} and up to  date it has been determined by the distance of the adjacent {H\sc ii} region/molecular cloud complex W3.  

HB3, was primarily detected in radio wavelengths by \cite{Brown1953} during a 158.5 MHz survey for radio sources in the Milky Way.  Further radio observations (\textit{e.g.} \citealt{ Williams1966, Caswell1967})  revealed the SNR nature of this radio source distinguishing it from the adjacent {H\sc ii} region W3.  In the radio band the remnant reveals a shell like morphology while its radio spectral index has been determined to be between – 0.64 and – 0.56 \citep{Landecker1987, Fesen1995, Reich2003, Tian2005, Green2007, Shi2008}. One pulsar discovered close to the SNR's boundaries seems not to be a part of it due to its large characteristic age \citep{Lorimer1998} and six masers detected in the line-of-sight of HB3 are most probably associated with the adjacent  W3 complex and not with the SNR \citep{Koralesky1998}.

Optical emission associated with HB3 was first found in the western portion of the remnant where a few filamentary structures were discovered \citep{ vandenBergh1973, DOdorico1977}.  Low dispersion spectra of these filaments exposed the shock heated origin of this emission and revealed a filament’s density of $ \le$ 100~cm$^{-3}$ and velocity of 35 – 40 $\rm  km~s^{-1}$  \citep[][respectively]{ DOdorico1977,Lozinskaya1980}. \cite{Fesen1983} and \cite{Fesen1995} presented new  $\rm H\alpha$ images of HB3 covering the western two-thirds of the remnant. The optical morphology of the remnant was found to be strongly spatially correlated with the radio band. Optical follow-up spectroscopy on the western limb filaments was also performed by \cite{Fesen1995}, indicating a shock velocity of $ \le 100~ \rm  km~s^{-1}  $, an  electron density of $ \le 150~ \rm cm^{-3}$ and an E(B-V)~=~$0.71 \pm 0.04$. 

The large size of HB3 in conjunction with the strong radio to optical correlation and the multi-shell appearance of the remnant, indicate a SNR of advanced age that most likely has passed beyond the adiabatic phases of evolution. However, the remnant is characterised by a centrally peaked thermal X-ray emission  \citep{Venkatesan1984, Leahy1985, Rho1998, Lazendic2006}. Such an evidence   cannot be explained by the canonical theory for SNR evolution, as in mature SNRs as HB3,  the reverse shock is expected to have been ceased long ago.  Thus, the remnant's interior is expected to be characterised by a cold and low density gas. For this reason, \citet{Rho1998}  classified HB3 as a mixed morphology (or thermal composite) SNR (MMSNR). X-ray spectroscopy revealed that the central emitting gas is close to ionization equilibrium characterised by a single temperature  \citep{Lazendic2006}. The element abundances of the X-ray emitting plasma are still ambiguous. \citet{Lazendic2006}  suggested that either it is characterised by enhanced abundances of O, Ne, Mg or it has marginally enhanced abundances of Mg and under-abundant Fe. 

Extended $\gamma$-ray emission is also discovered in the vicinity of the remnant \citep{Tsygankov2016,Katagiri2016}. This emission has been correlated to HB3 and it has been attributed to the decay of $\pi$\textsuperscript{0} produced by the interaction of  hadrons -accelerated at the remnant's shock waves- with the surrounding interstellar gas \citep{Katagiri2016}.

Regarding the environment of the remnant, HB3 has been found to evolve within and a rather complex medium where its eastern part is adjacent to the H{\sc ii} region / molecular complex W3 \citep[e.g.][]{Digel1996}.  \cite{Routledge1991} presented  radio observations on the 21~cm H{\sc i} emission covering the whole region of the remnants and $\rm ^{12} CO$ spectra in the southern portion of HB3. A shell shaped structure superimposed on HB3 was found indicating an atomic  neutral gas acceleration by the SNR blast wave. In addition, a bright $\rm ^{12} CO$ emission near $-$43~ km~s$^{-1}$ was detected overlapping with the radio emission of HB3 and they interpreted this evidence as a SNR interaction with molecular gas, in essence with W3. This evidence was verified by multimeter observations of CO lines, conducted by  \citet{Zhou2016} which revealed a substantial amount of molecular gas around $-$43~km~s$^{-1}$, being morphologically and dynamically  correlated with HB3. Finally, shocked molecular hydrogen ($\rm H_2$) and broad CO was recently detected in the near/mid infrared and millimeter bands, respectively, whose morphological and kinematic properties provide an additional evidence on the interaction of the SNR with the surrounding clouds \citep{Rho2021}. 
 
The origin and age of HB3   are still not well-determined. \citet{Lazendic2006} assuming that the SNR is in the adiabatic phase and adopting an evaporating clouds model for MMSNRs, estimated the SN explosion energy to be $7-20 \times 10^{50}$~erg and the age of the SNR between 25$-$30~kyrs. On the other hand, \citet{Zhou2016} considered that the remnant of HB3 has passed into the radiative phase and came to the conclusion that it is approximately $ 21 \pm 2 $~kyrs old and the explosion energy $E= \left(1.6 \pm 0.9\right) \times 10^{51}$~erg. Regarding the progenitor mass, \citet{Zhou2016} estimated a lower limit of 28~$\rm M_{\odot}$ by applying the linear relationship between the size of the wind-blown and the progenitor mass suggested by \citet{Chen2013}.

In this paper, we explore the optical emission of the SNR G132.7$+$1.3 in its full extent. In particular, \hnii, \sulfur\ and \oiii\ emission line images are presented, which show, for the first time, the complete structure of this large SNR in the optical band. Moreover, we present optical higher resolution imaging and long-slit spectroscopy at multiple areas of the SNR. HB3's optical emission is also compared with the radio, X--ray, $\gamma-$ray and molecular emission. In Section 2, we describe our imaging and spectroscopic observations, with the results and multi-wavelength comparison presented in Section 3. We discuss the SNR's properties, origin and evolution in Section 4, while in Section 5, we summarize the conclusions acquired from this work.

\section{Observations}

A brief summary and log of our observations are given in Table~\ref{table1}. In the following sections, we describe the observations of HB3 in more detail. 
\subsection{Imaging}

\begin{table*}  
\caption[]{Imaging and Spectral log. }  
\label{table1}
\begin{minipage}{17cm}
\begin{tabular}{lcccll}  
\noalign{\smallskip}  
\hline  
\multicolumn{6}{c}{WIDE- FIELD IMAGING} \\  
\hline
Filter & $\lambda_{\rm c}$ & $\Delta \lambda$ & Exposure time & \multicolumn{2}{c}{Date} \\
  & (\AA) & (\AA) & (sec) & &  \\
\hline
\HNII & 6570 & 75 & 2400 (7)\footnote{Numbers in parentheses represent the number of individual frames.}[2]\footnote{Numbers in brackets represent the number of different fields in each filter.} & 2009 Aug 19-20,22,25 & 2010 Aug 29-30 \\
\HNII & 6582 & 80 & 600 (22) [2] & 2019 Jul 27-29, Sep 30 & 2019 Oct 1-2\\
\OIII & 5010 & 28 & 2400 (1)[1] & 2009 Aug 23 &\\
\OIII & 5010 & 28 & 900 (25) [2] & 2018 Aug 7-8, 11, 17 & 2019 Sep 5, 15\\
\SII & 6720 & 27 & 2400 (5)[2] & 2009 Aug 20-21 & 2010 Aug 29-30\\
\SII & 6720 & 32 & 600 (17) [2] & 2019 Jul 27-29, Sep 30 & 2019 Oct 1-2\\
Continuum blue & 5470 & 230 & 180 [1] & 2009 Aug 23 & \\
Johnson V & 5380 & 980 & 20 (20) [2] & 2019 Jul 27-29, Sep 30 & 2019 Oct 1-2\\
Continuum red & 6096 & 134 & 180 [2] & 2009 Aug 19-22,25 & 2010 Aug 29-30 \\
SDSS-r & 6214 & 1290 & 20 (20) [2] & 2019 Jul 27-29, Sep 30 & 2019 Oct 1-2\\
\hline
\multicolumn{6}{c}{HIGHER-RESOLUTION IMAGING} \\  
\hline
Filter & $\lambda_{\rm c}$ & $\Delta \lambda$ & Exposure time & Area & Date \\
  & (\AA) & (\AA) & (sec)& &  \\
\hline
\ha$+$[N {\sc ii}] 6584 \AA & 6578 & 40 & 1800 & S1, S2, S3, S4, S5, S6 & 2017 Oct 18-19 \\
\SII & 6727  & 40 & 1800 & S1, S2, S3, S4, S5, S6 & 2017 Oct 18-19 \\
\OIII & 5011 & 30 & 1800 & S1, S2, S5, S6             & 2017 Oct 18-19\\
\OIII & 5010 & 28 & 600 (11) & S1, S2, S3 & 2021 July 13-15\\
\hline
\multicolumn{6}{c}{SPECTROSCOPY\footnote{The exposure time for all spectra is 3600 s.}} \\  
\hline  
Position & \multicolumn{2}{c}{Slit centers} & Offset\footnote{Spatial offset from the slit center in arcsec: N($=$North), S($=$South).} &  Aperture
length\footnote{Aperture lengths for each area in arcsec.} &  Date  \\
 & $\alpha$ & $\delta$ & &  &  \\
 & (h m s) & (\degr\ \arcmin\ \arcsec) & (arcsec) & (arcsec) &  \\  
\hline
Slit 1a & 02 11 26.4 & 62 44 42.1 & 96 S & 86 & 2009 Aug 22  \\
Slit 1b & 02 11 26.4 & 62 44 42.1 & 190 S & 29 & 2009 Aug 22  \\
Slit 2a & 02 13 16.9 & 62 56 47.9 & 138 S & 41 & 2010 Sep 10  \\
Slit 2b & 02 13 16.9 & 62 56 47.9 & 89 S & 10 & 2010 Sep 10 \\
Slit 2c & 02 13 16.9 & 62 56 47.9 & 36 S & 14 & 2010 Sep 10  \\
Slit 3a & 02 13 33.6 & 62 48 55.9 & 26 N & 37 & 2010 Sep 09 \\
Slit 3b & 02 13 33.6 & 62 48 55.9 & 95 N & 27 & 2010 Sep 09 \\
Slit 4a & 02 12 40.7 & 62 21 02.6 & 170 N & 63 & 2009 Aug 20 \\
Slit 5a & 02 22 22.7 & 62 50 10.7 & 118 S & 80 & 2009 Aug 22 \\
Slit 5b & 02 22 22.7 & 62 50 10.7 & 41 N & 117 & 2009 Aug 22 \\
Slit 5c & 02 22 21.7 & 62 50 52.2 & 157 S & 96 & 2010 Sep 08 \\
Slit 5d & 02 22 21.7 & 62 50 52.2 & 13 S & 61 & 2010 Sep 08 \\
Slit 6a & 02 23 38.6 & 62 44 57.7 & 190 S & 29 & 2010 Sep 03 \\
Slit 6b & 02 23 38.6 & 62 44 57.7 & 18 N & 33 & 2010 Sep 03 \\
Slit 6c & 02 23 38.0 & 62 44 53.6 & 187 S & 23 & 2010 Sep 07 \\
Slit 6d & 02 23 38.0 & 62 44 53.6 & 24 N & 31 & 2010 Sep 07 \\
Slit 7a & 02 19 54.0 & 62 23 57.6 & 12 S & 80 & 2009 Aug 23 \\
Slit 7b & 02 19 54.0 & 62 23 57.6 & 103 N & 8 & 2009 Aug 23 \\
Slit 8a & 02 21 15.8 & 62 23 58.0 & 128 S & 43 & 2010 Sep 05 \\
Slit 8b & 02 21 15.8 & 62 23 58.0 & 32 S & 18 & 2010 Sep 05 \\
\hline  
\end{tabular}
\end{minipage}
\end{table*}  
\begin{sidewaysfigure*}
\centering
\vspace*{+19cm}
\includegraphics[scale=0.63]{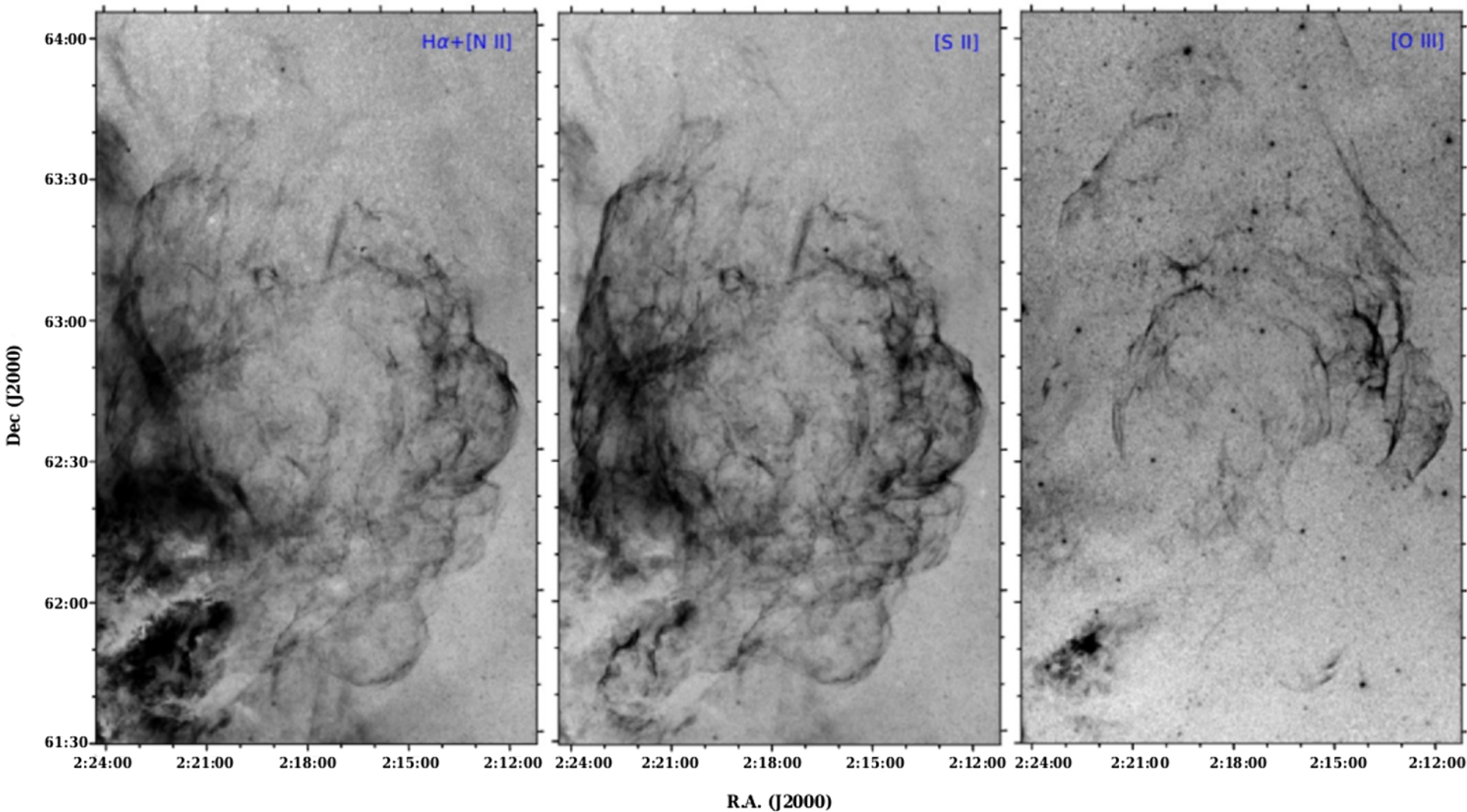}
\caption{The continuum--subtracted mosaic of G132.7$+$1.3 in the light of
\hnii, \sii\ and \oiii\ emission. The image centre is at R.A. 02$^{\rm h}$18$^{\rm m}$00$^{\rm s}$; Dec +62$\degr$45$\arcmin$00$\arcsec$ and its scale is $\sim$3$\arcsec$ pixel$^{-1}$.}
\label{fig1}
\end{sidewaysfigure*}

\begin{figure*}
\centering
\includegraphics[scale=0.9]{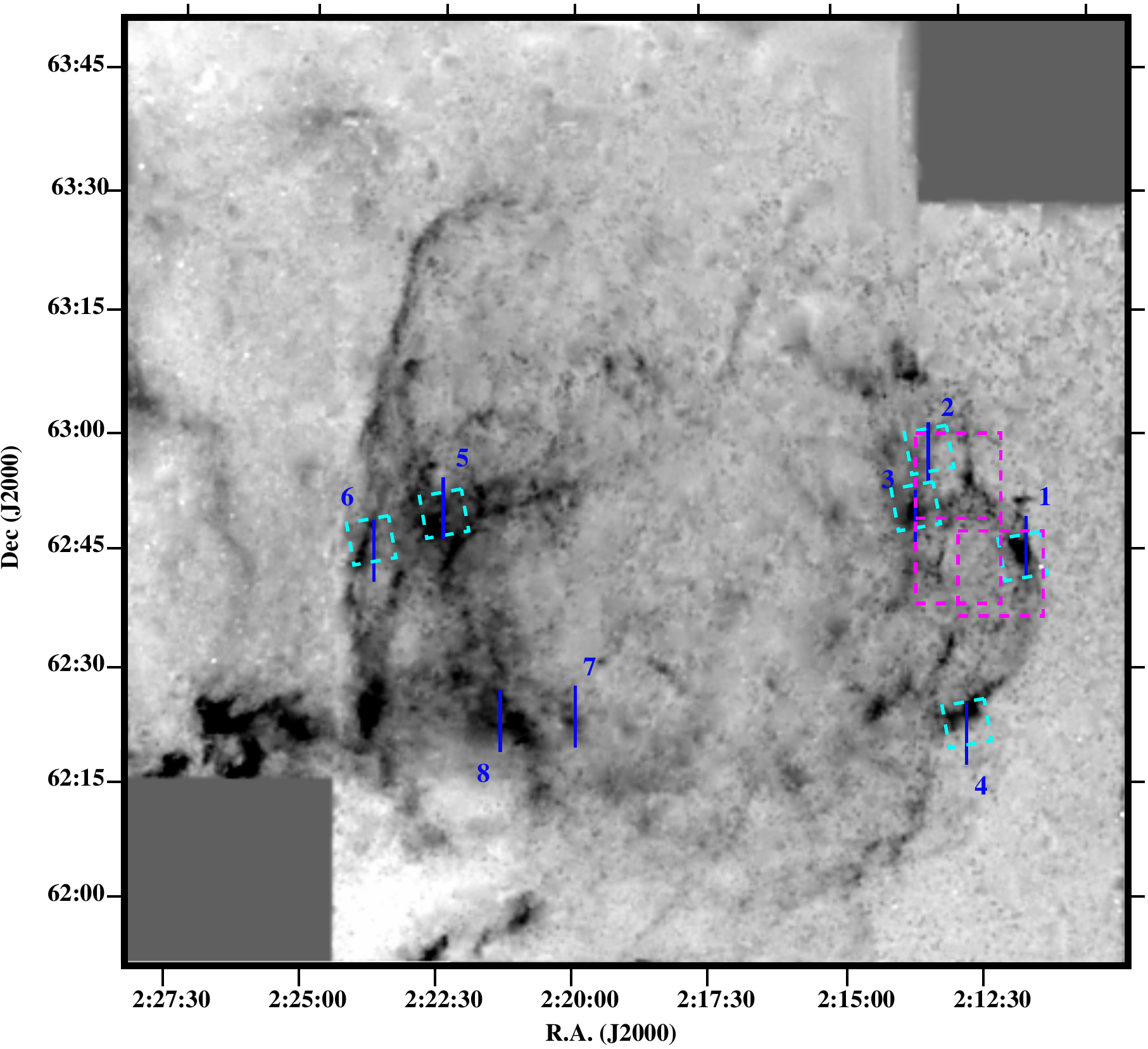}
\caption{The continuum--subtracted image of G132.7$+$1.3 in \sii\ emission. The blue lines indicate the positions of the slits, while the cyan  and magenta rectangles indicate the position of the higher resolution images shown in Fig.~\ref{fig4a}. Shadings run linearly from 0 to 39$\times$\flux\ (the maximum measured \sii\ flux on the SNR is 27$\times$\flux).}
\label{fig2}
\end{figure*}

\begin{figure*}
\centering
\includegraphics[scale=0.48]{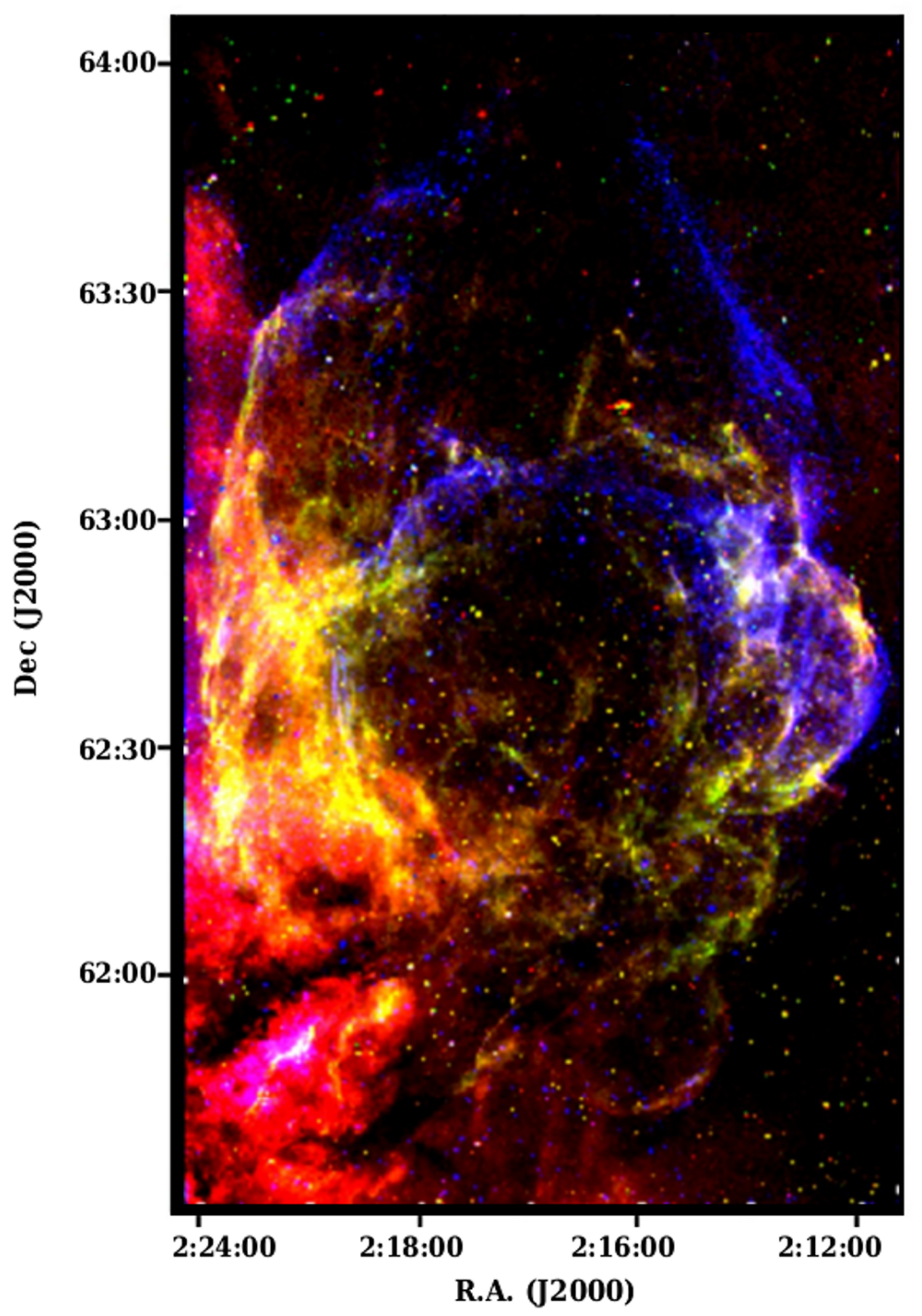}
\caption{The continuum--subtracted RGB image of G132.7$+$1.3, combining the \hnii\ (\textcolor{red}{red}), \sulfur\ (\textcolor{green}{green}) and \oiii\ (\textcolor{blue}{blue}) emission line images. The yellow colour produced in regions where both \hnii\ and \sulfur\ exist.}
\label{fig3}
\end{figure*}

\subsubsection{Wide-field imaging}

The wide-field imagery of HB3 was obtained with the 0.3 m Schmidt-Cassegrain (f/3.2) telescope at Skinakas Observatory, Crete, Greece in two different runs. The first was on the nights from 19$^{th}$ to 23$^{rd}$ and 25$^{th}$ of August 2009, as well as, on the 29$^{th}$ and 30$^{th}$ of August 2010. The camera used, was a $2048\times2048$ Andor DW436 CCD, which has a pixel size of 13.5~$\mu$m that results in a $101\arcmin\times101\arcmin$ field of view and an image scale of 2\arcsec.96 pixel$^{-1}$. In this run, the remnant was observed with the \hnii, \sulfur\ and \oiii\ filters with exposure times of 2400~s each and of 180~s for the continuum blue and red filters. Multiple exposures were taken in the first two filters, resulting to a total observing time of 16800~s for the \hnii\ and 12000~s for the \sii. The second run was on the nights 7$^{th}$, 8$^{th}$, 11$^{th}$ and 17$^{th}$ of August and on the 5$^{th}$ and 15$^{th}$ of September 2018, as well as, on the 27$^{th}$ -- 29$^{th}$ of July, 30$^{th}$ of September and 1$^{st}$ -- 2$^{nd}$ October 2019. The camera was the same with the first run, resulting to the same FOV and scaling. The filters used in this run were a new set of \hnii\ and \sii, the same \oiii, all with sets of exposure times of 600 s and the continuum SDSS-r and V with exposure time of 20 s (see Table~\ref{table1} for details). As in the first run, multiple exposures were taken for all filters, resulting to a total observing time of 13200 s for the \hnii, 10200 s for the \sii, 15000 s for the \oiii\ and 400 s for the continuum filters. During all the observations, the “seeing” varied between 0\arcsec.8 and 1\arcsec.2. 

For the data reduction the IRAF and Montage packages were used and all frames were bias-subtracted and flat-field corrected using a series of twilight flat-fields, and they were all airmass extinction corrected. For the absolute flux calibration the spectrophotometric standard stars HR 7596, HR 7950, HR 8634, HR~9087, and HR 0718 \citep{Hamuy1992, Hamuy1994} were used. The Hubble Space Telescope (HST) Guide Star Catalogue \citep{Lasker1999} was used for the calculation of the astrometric solution for all data frames and all the equatorial coordinates quoted in this
work refer to epoch 2000. The \hnii, \sulfur\ and \oiii\ continuum subtracted mosaic images are shown in Fig.~\ref{fig1}, while in Fig. \ref{fig2}, we present the flux-calibrated \sii\ image. In the latter image the blue lines indicate the positions of the slits, and the cyan  and magenta rectangles indicate the position of the higher resolution images shown in Fig.~\ref{fig4a}. For a better comparison between the three emission line images presented in Fig. \ref{fig1}, we created a colour RGB image shown in Fig. \ref{fig3}. In this figure, it is clear that most of the \hnii\ filaments coincide with the \sii\ ones, while there is a number of \oiii\ filaments which appear bright in this emission line and do not show any low ionization emission.

\subsubsection{Higher-resolution imaging}
\label{highres}
Follow-up, higher resolution images where obtained with the 2.3m (f/8) Aristarchos telescope at Helmos Observatory in Peloponnese, Greece, on 18$^{th}$ and 19$^{th}$ of October 2017 and with the 1.3m (f/7.64) Ritchey-Chretien telescope at Skinakas Observatory on 13$^{th}$ to 15$^{th}$ of July 2021. These images cover the regions of interest (hereafter areas S1 to S6) where spectroscopic observations were performed (centered at slit positions 1-6). In particular, S1--S6 were observed with the \hnii\ and \sii\ filters with Aristarchos telescope (cyan rectangles in Fig.~\ref{fig2}), while areas S1, S2, S5, S6 and S1, S2, S3 were observed with the \oiii\ filter with the Aristarchos (cyan rectangles in Fig.~\ref{fig2}) and Skinakas (magenta rectangles in Fig.~\ref{fig2}) telescopes, respectively.\\
The detector at Aristarchos telescope was a 2048$\times$2048, 13.5 $\mu$m pixels CCD, with a field of view of 5\arcmin.5$\times$5\arcmin.5 (0\arcsec.16 pixel$^{-1}$ in 2$\times$2 binning), while at Skinakas Observatory a 2048$\times$2048, 13.5~$\mu$m pixels Andor CCD was used which has a 9\arcmin.5$\times$9\arcmin.5 field of view and an image scale of 0\arcsec.28 pixel$^{-1}$. All information about the used \hnii, \sulfur\ and \oiii\ filters as well as the exposure times of all higher-resolution images are shown in \autoref{table1}. The data reduction was carried out using standard IRAF routine packages for the bias subtraction and flat-field correction, while the same catalogue of stars was used for the astrometric calculations. Seeing conditions were between 1\arcsec\ to 1\arcsec.5.\\
The higher-resolution images are shown in Figs.~\ref{fig4a}, with two of them (S1 and S6) being also presented in RGB colour images (Fig.~\ref{fig5}). Details of all imaging observations are given in Table~\ref{table1}.

\begin{sidewaysfigure*}
\centering
\vspace*{+19cm}
\includegraphics[scale=0.6]{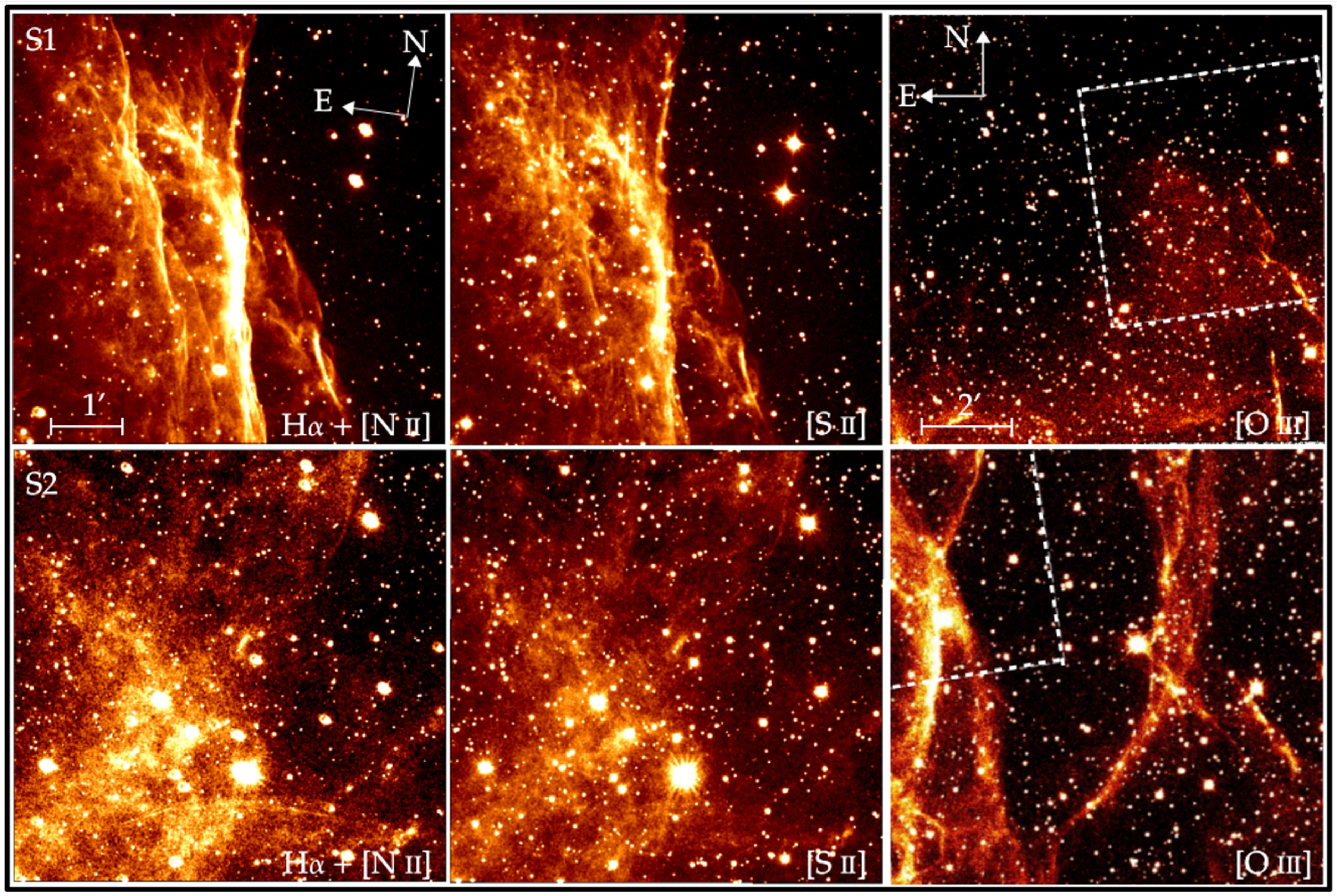}
\caption{Higher resolution images of G132.7$+$1.3 in
\hnii, \sii and \oiii\ emission of the regions where slit positions 1-6 were taken (see Fig. \ref{fig2}). The North-East orientation is the same for \hnii\ and \sii\ images and it can be seen on the top left image. The \oiii\ images have a larger FOV (Sect.~\ref{highres}), so the position of the \hnii, \sii\ images are indicated with a dash rectangle}. The faint ring which appears to the north-east of all bright stars is a ghost due to the filter.
\label{fig4a}
\end{sidewaysfigure*}

\begin{sidewaysfigure*}
\addtocounter{figure}{-1}
\centering
\vspace*{-19cm}
\includegraphics[scale=0.6]{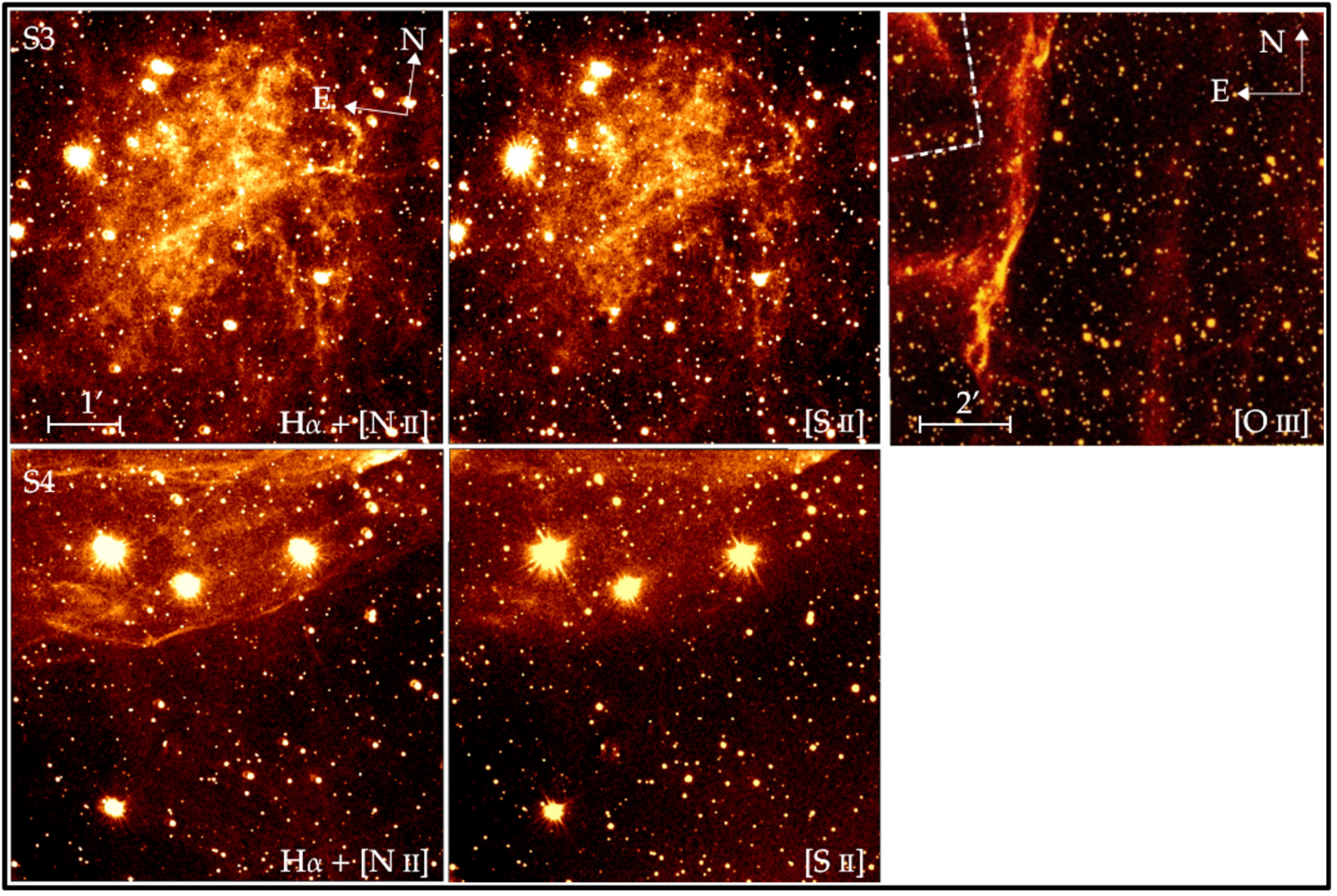}
\caption{Continued.}
\label{fig4b}
\end{sidewaysfigure*}

\begin{sidewaysfigure*}
\addtocounter{figure}{-1}
\centering
\vspace*{+19cm}
\includegraphics[scale=0.6]{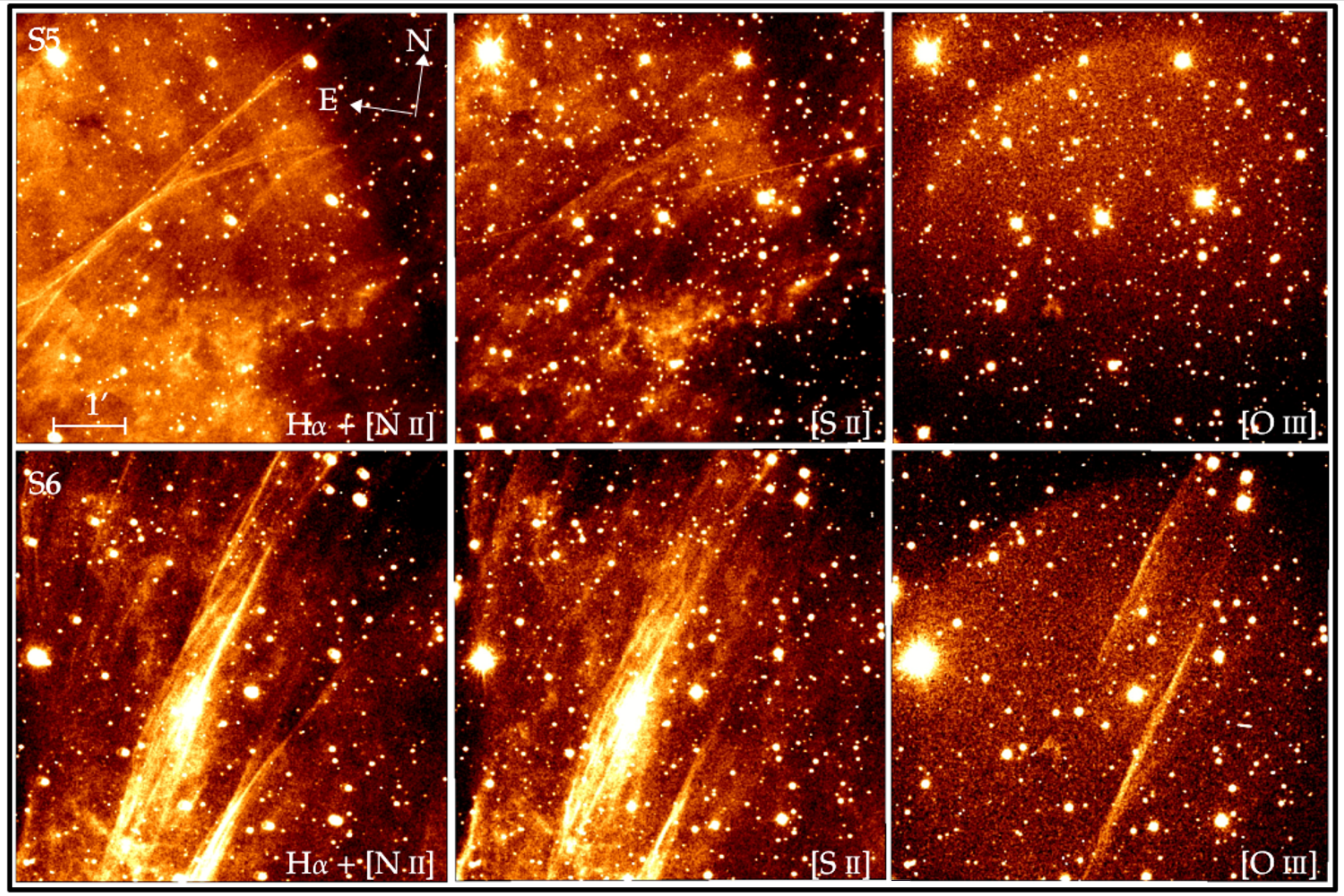}
\caption{Continued.}
\label{fig4c}
\end{sidewaysfigure*}

\begin{figure*}
\centering
\includegraphics[scale=0.44]{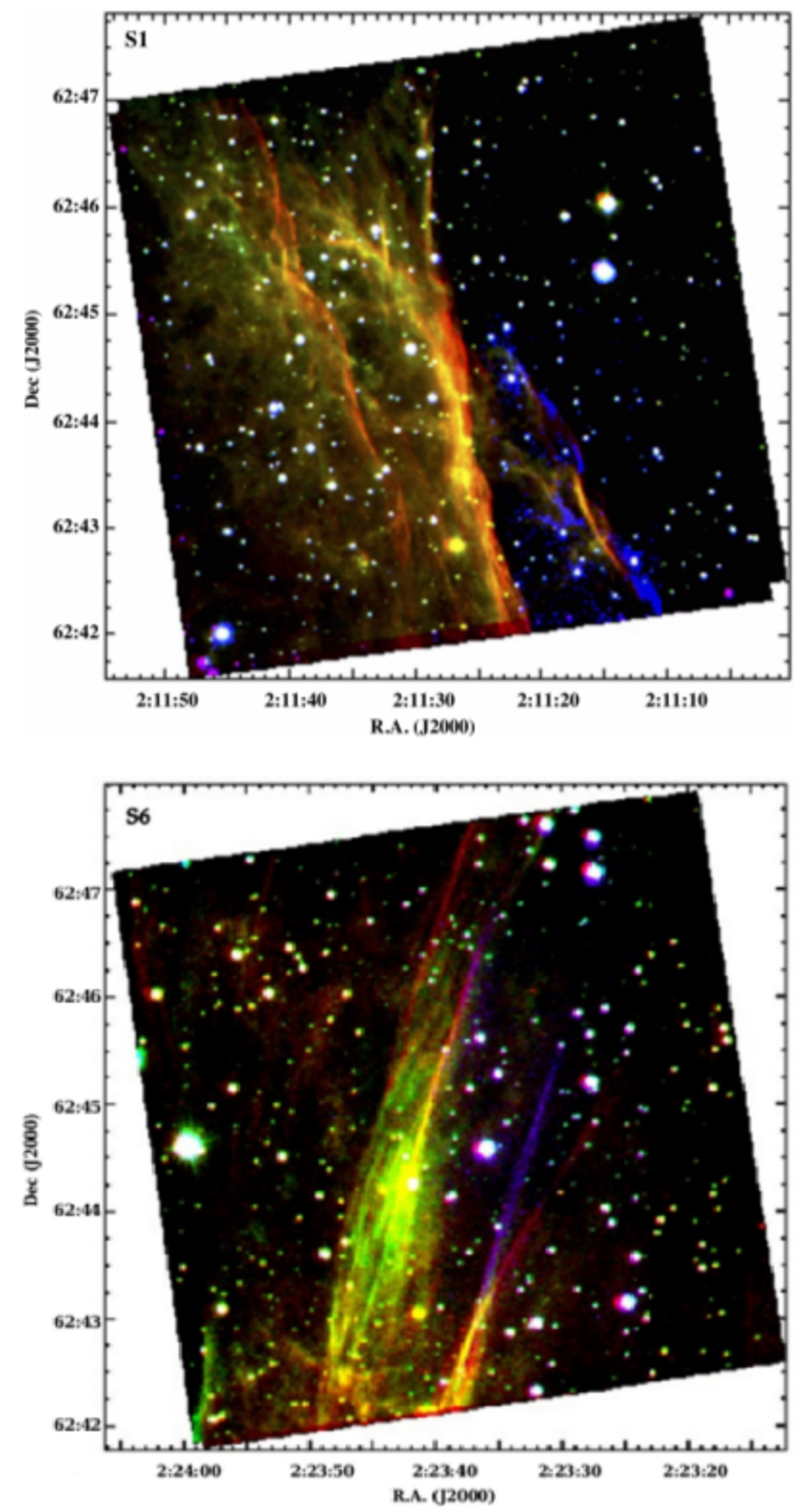}
\caption{Higher resolution RGB image of Areas S1 and S6, combining the \hnii\ (\textcolor{red}{red}), \sulfur\ (\textcolor{green}{green}) and \oiii\ (\textcolor{blue}{blue}) emission line images. The yellow colour produced in regions where both \hnii\ and \sulfur\ exist. North is to the top, East to the left.}
\label{fig5}
\end{figure*}

\subsection{Spectroscopy}
Low-dispersion, long-slit spectra where obtained with the 1.3 m Ritchey-Cretien telescope (f/7.7), at Skinakas Observatory,  on August 20, 22-23, 2009 and September 3, 5, 7-10, 2010. For these observations, the 1302 line \,mm$^{-1}$ grating was employed with the $2000\times800$ (13 $\mu$m) SITe CCD covering the wavelength range from 4750 to 6815\AA, while the exposure time for each data frame was 3600 s. The spectral resolution being $\sim$8~\AA \, and $\sim$11~\AA~for the red and blue part of the spectra, respectively. The slit has a width of 7.7\,arcsec and a length of 7.9\,arcmin and it was oriented in the north-south direction for all the positions. The coordinates of each slit are given in Table~\ref{table1}. For the data reduction the IRAF package was used, while the spectrophotometric standards stars HR 0718, HR 7596, HR 7950, HR 8634, and HR 9087 were used for the flux calibration of the spectra \citep{Hamuy1992, Hamuy1994}. 

\begin{figure*}
\centering
\includegraphics[width=\textwidth]{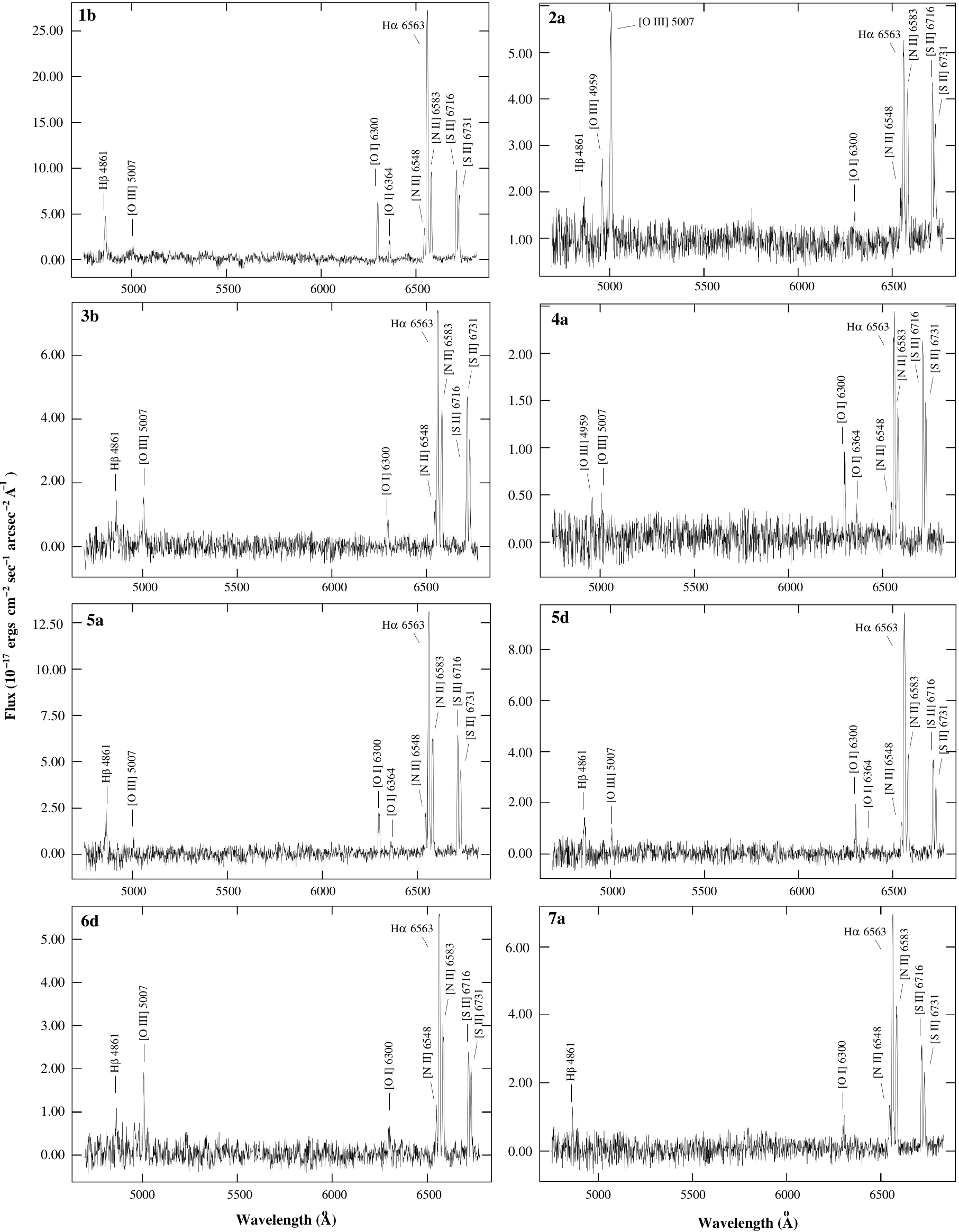}
\caption{Long--slit low resolution spectra from different positions of the observed area (see Table \ref{table1}). The small shift in the wavelength axis is due to the different observing season (slightly different position of the grating).
}
\label{fig8}
\end{figure*}

\begin{figure*}
\centering
\includegraphics[scale=0.45]{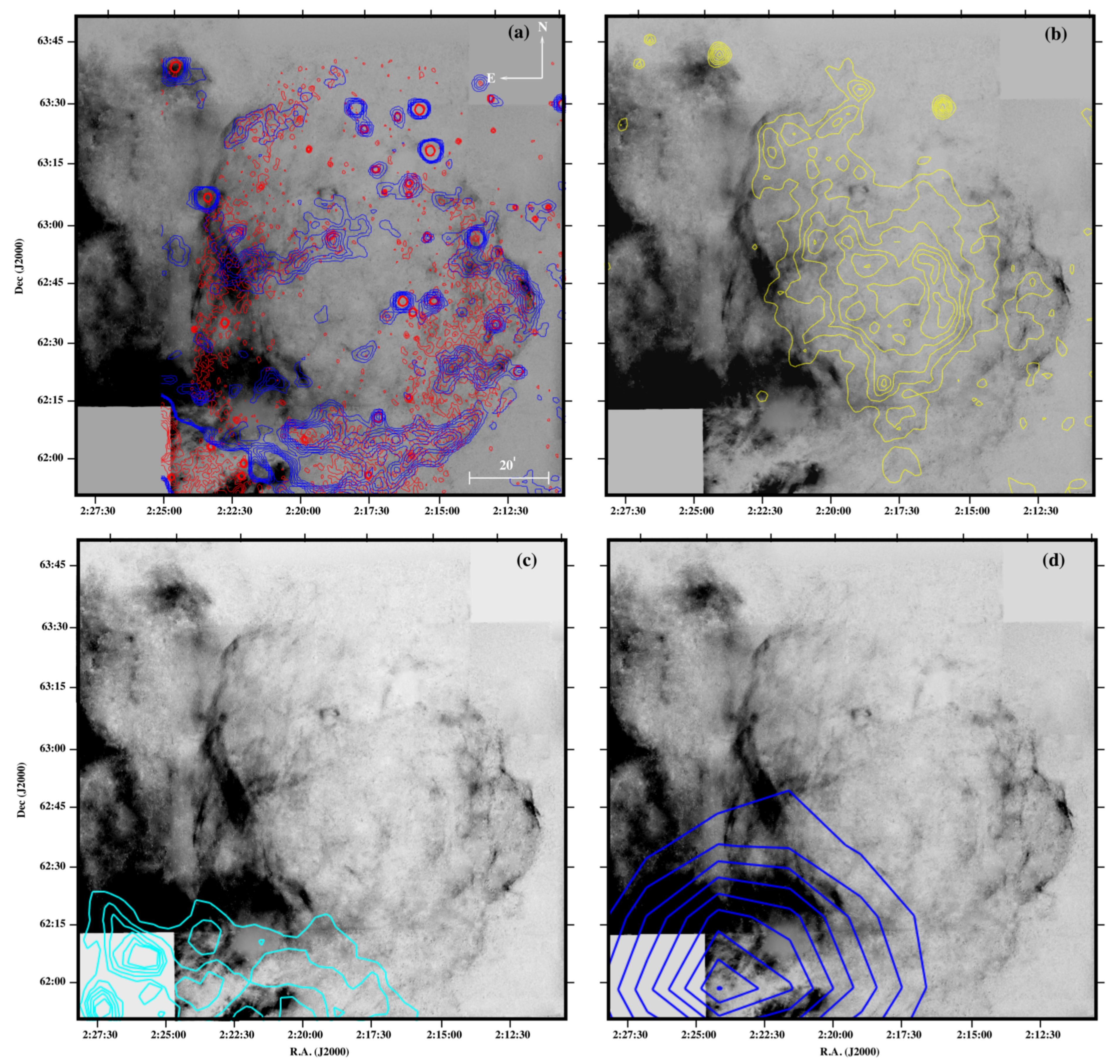}
\caption{The correlation between \hnii\ image and {\ bf(a)} the radio emission
at 4850 MHz (blue line) and 330 MHz (red line). The blue contours
scale linearly from 5$\times 10^{-3}$~Jy/beam to 0.03 Jy/beam, with
step 0.05 Jy/beam and the red from 7$\times 10^{-3}$~Jy/beam to
4.5$\times 10^{-2}$~Jy/beam, with step 0.05 Jy/beam. {\bf (b)} the X-ray emission of {\it ROSAT} PSPC (yellow line, \citealt{Lazendic2006}). The yellow contours scale linearly from 1$\times 10^{-5}$~counts arcmin$^{-2}$ s$^{-1}$ to 3$\times 10^{-4}$~counts arcmin$^{-2}$ s$^{-1}$, with step 2.8$\times 10^{-5}$~counts arcmin$^{-2}$ s$^{-1}$. {\bf (c)} the molecular emission of $^{12}$CO ($J=1-0$), $^{13}$CO ($J=1-0$) and C$^{18}$O ($J=1-0$) in the velocity range of $-$60 to $-$30 \vel\ (cyan line, \citealt{Zhou2016}). The cyan contours scale linearly from 0 to 13.5 K with step 5 K. {\bf (d)} the $\gamma-$ray emission of background-subtracted {\it FERMI} LAT in the $1-10$ GeV (blue line, \citealt{Katagiri2016}).The blue contours scale linearly from 0 to 22 counts pixel$^{-1}$ with step 5 counts pixel$^{-1}$.}
\label{fig9}
\end{figure*}

\section{Results}

\begin{table*}
\caption[]{Typically measured fluxes over the brightest filaments with median values over a 40\arcsec $\times$ 40\arcsec\ box. Fluxes are in units of \fluxa. }
\label{table2}
\begin{minipage}{16cm}
\begin{tabular}{llllllllll}
\hline
\noalign{\smallskip}
 & Area S1 & Area S2 & Area S3 & Area S4 & Area S5 & Area S6 & Area S7 & Area S8 & Area A\footnote{Position of center of Area A: $\alpha$= 02\textsuperscript{h} 23\textsuperscript{m} 32\textsuperscript{s}, $\delta$= + 62\degr\ 21\arcmin\ 37\arcsec\ .} \\
\hline
\hnii\  & 188 & 63 & 112 & 111 & 134 & 114 & 96 & 146 & 257\\
\hline
\sii\   & 54 & $<$ 15 & 34 & 42 & 39 & 44 & 17 & 30 & 38 \\
\hline
\oiii\footnote{3$\sigma$~upper limit.} & & & & & $<$ 16 & $<$ 13 & $<$ 16 & $<$ 16 & $<$ 20 \\
\hline
\sii/\ha\ & 0.58 & 0.47 & 0.60 & 0.74 & 0.58 & 0.78 & 0.35 & 0.41 & 0.30 \\
\hline
\end{tabular}
\end{minipage}
\end{table*}

\subsection{The optical emission line images}

The optical emission of HB3 is detected for the first time in its full extent. Both bright and faint, diffuse and filamentary structures spread along the entire remnant are revealed, while the {H\sc ii} region W3 is also noted at the southeast region of the remnant. The \sulfur\ emission of HB3, though fainter, displays a similar morphology to that of the \hnii\ emission-line image, tracing all the filamentary structures. The \oiii\ emission line, displays a different filamentary morphology compared to the former line images, indicating different shock velocities and physical conditions.

\hnii, \sulfur\: and \oiii\: fluxes are calculated from areas with 40x40~arcsec$^2$ dimensions over the brightest filaments where the low resolution spectra were taken (see Table \ref{table2}). The errors of the integrated fluxes are estimated between 10 and 20 percent. All selected slit areas but one (area~S7) and area A (where a known {H\sc ii} region exists) exhibit \sulfur/H$\alpha>$0.4, as expected for shock-heated regions.    

The most interesting regions lie in the west, south-west and east directions, where bright large filamentary and faint diffuse structures are
present (Figs. \ref{fig1}, \ref{fig4a}, between $\alpha \simeq$ 02\h19\m30\s, $\delta \simeq$
62\degr15\arcmin; $\alpha \simeq$ 02\h24\m, $\delta
\simeq$ 63\degr30\arcmin\ and $\alpha \simeq$ 02\h11\m10\s,
$\delta \simeq$ 62\degr23\arcmin; $\alpha \simeq$ 02\h14\m30\s,
$\delta \simeq$ 63\degr13\arcmin), which all are very well
correlated with the radio emission. The bright \hnii\ filaments cover
most of the emission found at radio wavelengths. In contrast to the above results, diffuse emission was mainly
detected in the south, centre and north with only a few small filamentary structures are present (i.e. those at $\alpha
\simeq$ 02\h18\m40\s, $\delta \simeq$ 63\degr09\arcmin10\arcsec and
$\alpha \simeq$ 02\h19\m15\s, $\delta \simeq$
63\degr09\arcmin20\arcsec\ ; $\sim$2\arcmin\ long). We also detected \sii\ emission where most of the \hnii\ emission
was found with filamentary bright structures and diffuse emission in the same regions (Figs. \ref{fig1}, \ref{fig2}, \ref{fig4b}). Both the \hnii\ and \sii\ higher--resolution images (Fig. \ref{fig4a} show clearly  filamentary and diffuse emission in all slit positions 1--6, which have similar structures. Furthermore, very thin (less than 2-3\arcsec\ wide), as well as wider ($\sim$10\arcsec) filaments are present, while the shock front region can be clearly seen (i.e. in S1 and S4 areas).
Note that the detected \oiii\ emission (Figs. \ref{fig1}, \ref{fig4a}) appears
more filamentary and less diffuse than in the \hnii\ and \sii\ images. In
Table~\ref{table2} upper limit fluxes are listed for the \oiii~$\lambda$5007~line. Significant
differences between the \hnii\ and \oiii\ images are present for many
of the filaments. In particular, most of the bright \hnii\ and \sii\ complex filamentary structures to the east, south--east are not present to the \oiii\ image, while there are other bright filaments only to the \oiii\ emission. However, some thin filaments (of the order of 8-10\arcmin\ long), coincide with the \hnii\ and \sii\ emission (see Fig. \ref{fig1}). These are at slit positions 5 and 6 and to the north at $\alpha \simeq$ 02\h22\m55\s, $\delta \simeq$
63\degr25\arcmin25\arcsec\ and $\alpha \simeq$ 02\h22\m07\s, $\delta \simeq$
63\degr30\arcmin00\arcsec. Further to the east, there is a number of bright
\oiii\ filaments which are co-spatial with the \hnii\ structures, but still it is clear that the bright \oiii\ filaments do not coincide with those in the \hnii\ and \sii\ emission. Interestingly, there are many thin \oiii\ filamentary structures to the central region of HB3 (i.e. at $\alpha \simeq$ 02\h20\m16\s, $\delta \simeq$
62\degr59\arcmin00\arcsec, $\alpha \simeq$ 02\h19\m50\s, $\delta \simeq$
62\degr53\arcmin00\arcsec\ and $\alpha \simeq$ 02\h14\m55\s, $\delta \simeq$
62\degr49\arcmin30\arcsec), which do not show any emission in \hnii\ and \sii.
Taking into account all the emission line images presented in Fig. \ref{fig1}, and the RGB image presented in Fig. \ref{fig3}, it seems that HB3 traces an axi-symmetric bi-lobal structure with respect to its centre. This is an interesting approach and should be investigated further in the future when kinematical data will be available.
In Fig.~\ref{fig5}, we present an RGB image of slit positions 1 and 6 regions, where the \hnii\ (red), \sii\ (green) and \oiii\ (blue) higher--resolution images are combined. In the S1 image, the \hnii\ emission of the small thin filaments (i.e. at $\alpha \simeq$ 02\h11\m15\s, $\delta \simeq$
62\degr43\arcmin00\arcsec)  in front of the large ones (i.e. at $\alpha \simeq$ 02\h11\m30\s, $\delta \simeq$
62\degr44\arcmin00\arcsec) coincide with that of the \oiii\ and lie immediately ahead of the diffuse \oiii\ emission (similar to SNR G 107.0$+$9.0; \citealt{Fesen2020}). Unfortunately, there is no \oiii\ emission where the bright \hnii\ filaments are, so we cannot have a complete and clear view of the whole region.

\subsection{The optical spectra}

Deep low-resolution long-slit spectra were obtained for the brightest filaments of the remnant. The exact position of the silts is illustrated in Fig. \ref{fig2} and the coordinates of their centres are listed in Table~\ref{table1}. The relative line fluxes as well as the signal-to-noise ratios for multiple apertures free of background stars are computed and listed in Table~\ref{table3}. The absolute H$\alpha$ fluxes are also determined for all the apertures and vary from 5 to 192 $\times$ \fluxa. The lengths of these apertures and their offsets from the centre of the slits are also given in Table \ref{table1}. For the extraction of the background emission, the areas were chosen north or south of the centre of the slits as being free of field stars and depending on the location of the filaments. Fig.~\ref{fig8} presents as representative examples the 1D spectra of six apertures. 

Our slit~1 covers the western filament, which has also been studied by \cite{DOdorico1977} and \cite{Fesen1995}, and it shows a good agreement despite the different position and orientation. In particular, we determine an \sii/H$\alpha$ ratio equals to 0.8 and 0.56 for the apertures 1a and 1b, respectively, very close to the values of 0.72 reported by \cite{Fesen1995} and 0.62 from \cite{DOdorico1977}. The \siirat~line ratio is calculated 1.32$\pm$0.06 and 1.39$\pm$0.05, close to the value of 1.5$\pm$0.2 found by \cite{Fesen1995}, and even closer to the value of 1.44$\pm$0.01 from \cite{DOdorico1977}. A consistency between the studies is also found in the interstellar reddening (0.71$\pm$0.04; \citealt{Fesen1995} and a median of 0.76 in this study). The \OI~$\lambda$6300 line, a diagnostic for shock heated gas in SNRs \citep{Kopsacheili2020}, is also detected in our spectrum of slit~1. However, it has to be noted that the \NI~$\lambda$5200 line which is reported by \cite{Fesen1995} is not detected in our spectrum, while the \oiii~$\lambda$5007/\hb\ ratio is three times higher in our data. Fig.~\ref{fig5} displays the RGB images of the western (area~S1) and eastern (area-S6) filaments. At the former (S1), there are interestingly two main filaments bright in \hnii\  and \oiii~$\lambda$5007 with an offset of a few arcsec. The second displays some filaments in \hnii\ emission. This explains the difference in the \oiii~$\lambda$5007/\hb\ ratio between our result and the one from \cite{Fesen1995} as their slit do not cover the \oiii~dominated filament resulting in lower \oiii~$\lambda$5007/\hb\ ratio.

The common criterion for shock heated gas in SNRs, the \sii /H$\alpha$ line ratio, is higher than 0.4 for all the filaments observed, while the [N {\sc ii}]/H$\alpha$ ratio varies from 0.5 to 1 (see Table \ref{table3}). Both ratios verify the SNR nature of HB3 (see also \citealt{Fesen1983,Fesen1995}). All the filaments are distributed in the locus of typical shock-heated SNRs well separated from the photo-ionized nebulae.

Note that all slits/apertures but two (2a and 2b) exhibit a weak or even not detectable \oiii~5007\AA~line. The log(\oiii/H$\beta$) ratio is lower than 0.5 which indicates shocks with complete recombination lines (\citealt{cox1985}; \citealt{Hartigan1987}) and velocities between 80-100~\vel\ (\citealt{Raymond1988}; \citealt{Osterbrock2006}). The absence of the \oiii~5007\AA~emission line in some filaments implies the presence of shock with even lower velocities ($<$80~km~s\textsuperscript{-1}), and this result is consistent with the observed \oi/H$\alpha$ ratios. On the other hand, high \oiii/H$\beta$ ratio ($>$0.8) is found in slit~2 (apertures a and b, see Table \ref{table3}). Such a strong \oiii\ line indicates shock velocities higher than 100~km~s\textsuperscript{-1}. Furthermore, having a look at Figs. \ref{fig1} -- \ref{fig3}, it can be clearly seen that there are many filaments with strong \oiii\ emission, which are not covered by the available slits, where even higher shock velocities ($>$120 km s$^{-1}$) are expected to be found.

The electron density of the filaments is lower than 240 cm\textsuperscript{-3} obtained using the \sulfur~diagnostic lines and the \textit{temden} STSDAS/IRAF routine. Due to the absence of the typical \oxygen~$\lambda$4363 and \nitrogen~$\lambda$5755 diagnostic lines (based on \citealt{Osterbrock2006}), we are not able to compute the electron temperature of the gas and a value of 10\textsuperscript{4}~K is considered for all the filaments.

Finally, for all filaments where the H$\beta$ line is detected, the interstellar extinction coefficient c(H$\beta$) is computed and varies from 0.62 ($\pm$ 0.22) up to 1.14 ($\pm$ 0.36) with a mean value of 0.9 or an A\textsubscript{v} from 1.31 ($\pm$ 0.48) to 2.32 ($\pm$ 0.76)~, as presented in Table \ref{table3}. 

\begin{table*}
\caption[]{Relative line fluxes. The emission line ratios \sulfur/\ha\ , F(6716)/F(6731) and \nitrogen/\ha\ are calculated using either the values corrected for interstellar extinction (when \hbeta\ line is available) or the values uncorrected for interstellar extinction.The errors of the emission line ratios, c(\hbeta) and E$_{\rm B-V}$ are calculated through standard error propagation.}
\label{table3}
\begin{minipage}{17cm}
\begin{tabular}{lccccccccccccccc}
\hline
\noalign{\smallskip}
 & \multicolumn{3}{c}{Slit 1a} & \multicolumn{3}{c}{Slit 1b} 
& \multicolumn{3}{c}{Slit 2a} & \multicolumn{3}{c}{Slit 2b} &
\multicolumn{3}{c}{Slit 2c} \\ 
Line (\AA) & F\footnote{Observed fluxes normalized to F(H$\alpha$)=100 and
uncorrected for interstellar extinction.} & I\footnote{Observed fluxes normalized to F(H$\alpha$)=100 and
corrected for interstellar extinction.}  & S/N\footnote{Numbers represent the signal to noise ratio of the quoted fluxes.} & F & I & S/N
& F & I & S/N & F & I & S/N & F & I & S/N \\
\hline
\hbeta\ 4861 & 20 & 35 & 6 & 18 & 35 & 9 & 14 & 35 & 3 & 17 & 35 & 3 & $-$  & $-$  &  $-$  \\
\oxygen\ 4959 & $-$ & $-$ & $-$ & $-$ & $-$ & $-$ & 37 & 88 & 5 & 45 & 87 & 5 & 17 & $-$ & 3 \\
\oxygen\ 5007 & 6 & 11 & 3 & 3 & 6 & 4 & 127 & 281 & 13 & 152 & 288 & 16 & 63 & $-$  & 4 \\
\nitrogen\ 5755 & $-$ & $-$ & $-$ & $-$ & $-$ & $-$ & 3 & 5 & 3 & $-$ & $-$ & $-$ & $-$ & $-$ & $-$ \\
\oi\ 6300 & 18 & 20 & 9 & 21 & 24 & 14 & 13 & 15 & 4 & 6 & 7 & 4 & 14 & $-$  & 3 \\
\oi\ 6364 & $-$ & $-$ & $-$ & 7 & 8 & 6 &  $-$ & $-$ & $-$ & $-$ & $-$ & $-$ & $-$ & $-$ &  $-$ \\
\nitrogen\ 6548 & 16 & 17 & 10 & 11 & 11 & 10 & 25 & 26 & 7 & 30 & 32 & 5 & 30 & $-$ & 4 \\
\ha\ 6563 & 100 & 100 & 62 & 100 & 100 &    92 & 100 & 100 & 26 & 100 & 100 & 16 & 100 & $-$ & 14 \\
\nitrogen\ 6584 & 49 & 49 & 29 & 33 & 33 & 32 & 79 & 79 & 18 & 85 & 85 & 13 & 86 & $-$ & 13 \\
\sulfur\ 6716 & 46 & 46 & 29 & 33 & 33 & 32 & 75 & 74 & 18 & 64 & 64 & 13 & 65 & $-$ & 13 \\
\sulfur\ 6731 & 35 & 35 & 21 & 24 & 23 & 26 & 63 & 62 & 15 & 56 & 55 & 10 & 43 & $-$ & 7 \\
\hline
Absolute \ha\ flux\footnote{In units of \fluxa.} & \multicolumn{3}{c}{191.5} &
\multicolumn{3}{c}{122.8} & \multicolumn{3}{c}{26.2} &
\multicolumn{3}{c}{4.9} & \multicolumn{3}{c}{5.1} \\

\sulfur/\ha\ & \multicolumn{3}{c}{0.80 $\pm$ 0.05} &
\multicolumn{3}{c}{0.56 $\pm$ 0.03} & \multicolumn{3}{c}{1.34 $\pm$ 0.12} & \multicolumn{3}{c}{1.19 $\pm$ 0.18} & \multicolumn{3}{c}{1.08 $\pm$ 0.20} \\
F(6716)/F(6731) & \multicolumn{3}{c}{1.32 $\pm$ 0.06} &
\multicolumn{3}{c}{1.39 $\pm$ 0.05} & \multicolumn{3}{c}{1.21 $\pm$ 0.08} &
\multicolumn{3}{c}{1.16 $\pm$ 0.12} & \multicolumn{3}{c}{1.50 $\pm$ 0.21} \\

\nitrogen/\ha\ & \multicolumn{3}{c}{0.65 $\pm$ 0.07} &
\multicolumn{3}{c}{0.44 $\pm$ 0.05} & \multicolumn{3}{c}{1.07 $\pm$ 0.17} & \multicolumn{3}{c}{1.16 $\pm$ 0.25} & \multicolumn{3}{c}{1.15 $\pm$ 0.31} \\

c(\hbeta)\footnote{The logarithmic extinction is derived by c =
1/0.348$\times$log((\ha/\hbeta)$_{\rm obs}$/2.85).} & \multicolumn{3}{c}{0.69 $\pm$ 0.14} &
\multicolumn{3}{c}{0.86 $\pm$ 0.06} & \multicolumn{3}{c}{1.14 $\pm$ 0.41} & \multicolumn{3}{c}{0.88 $\pm$ 0.43} & \multicolumn{3}{c}{$-$} \\
E$_{\rm B-V}$\footnote{The interstellar reddening was measured from the relation E$_{\rm B-V}$ $\approx$ 0.77c (\citealt{Osterbrock2006})} & \multicolumn{3}{c}{0.53 $\pm$ 0.11} &
\multicolumn{3}{c}{0.66 $\pm$ 0.05} & \multicolumn{3}{c}{0.88 $\pm$ 0.30} & \multicolumn{3}{c}{0.68 $\pm$ 0.33} & \multicolumn{3}{c}{$-$}\\
\hline
 & \multicolumn{3}{c}{Slit 3a} &
\multicolumn{3}{c}{Slit 3b} & \multicolumn{3}{c}{Slit 4a} &
\multicolumn{3}{c}{Slit 5a} & \multicolumn{3}{c}{Slit 5b} \\ 
Line (\AA) & F & I  & S/N & F & I & S/N & F & I & S/N & F & I & S/N & F
& I & S/N \\
\hline 
\hbeta\ 4861 & 21 & 35 & 4 & 17 & 35 & 5 & - & - & - & 16 & 35 & 6 & 15 & 35 & 6 \\
\oxygen\ 4959 & $-$ & $-$ & $-$ & $-$ & $-$ & $-$ & 5 & $-$ & 2 & $-$ & $-$ & $-$ & $-$ & $-$ & $-$ \\
\oxygen\ 5007 & 32 & 50 & 9 & 20 & 38 & 6 & 3 & $-$ & 3 & 4 & 8 & 3 & 1 & 2 & 2 \\
\oi\ 6300 & 14 & 15 & 6 & 10 & 11 & 5 & 30 & $-$ & 8 & 16 & 19 & 7 & 16 & 18 & 4 \\
\oi\ 6364 & $-$ & $-$ & $-$ & $-$ & $-$ & $-$ & 10 & $-$ & 3 & 5 & 6 & 3 & $-$ & $-$ & $-$ \\
\nitrogen\ 6548 & 20 & 21 & 7 & 16 & 17 & 7 & 18 & $-$ & 5 & 14 & 15 & 7 & 13 & 14 & 6 \\
\ha\ 6563 & 100 & 100 & 29 & 100 & 100 & 31 & 100 & $-$ & 28 & 100 & 100 & 38 & 100 & 100 & 35 \\
\nitrogen\ 6584 & 63 & 63 & 18 & 56 & 56 & 19 & 62 & $-$ & 16 & 49 & 49 & 19 & 41 & 41 & 16 \\
\sulfur\ 6716 & 70 & 69 & 21 & 59 & 59 & 19 & 89 & $-$ & 23 & 49 & 49 & 21 & 36 & 35 & 14 \\
\sulfur\ 6731 & 48 & 47 & 15 & 42 & 42 & 13 & 64 & $-$ & 17 & 35 & 34 & 16 & 24 & 24 & 12 \\
\hline 
Absolute \ha\ flux & \multicolumn{3}{c}{33.8} &
\multicolumn{3}{c}{30.3} & \multicolumn{3}{c}{23.0} &
\multicolumn{3}{c}{153.6} & \multicolumn{3}{c}{168.2} \\

\sulfur/\ha\ & \multicolumn{3}{c}{1.17 $\pm$ 0.11} &
\multicolumn{3}{c}{1.00 $\pm$ 0.09} & \multicolumn{3}{c}{1.53 $\pm$ 0.13} & \multicolumn{3}{c}{0.83 $\pm$ 0.07} & \multicolumn{3}{c}{0.59 $\pm$ 0.07} \\

F(6716)/F(6731) & \multicolumn{3}{c}{1.47 $\pm$ 0.10} &
\multicolumn{3}{c}{1.42 $\pm$ 0.09} & \multicolumn{3}{c}{1.38 $\pm$ 0.08} & \multicolumn{3}{c}{1.43 $\pm$ 0.09} & \multicolumn{3}{c}{1.47 $\pm$ 0.12} \\

\nitrogen/\ha\ &\multicolumn{3}{c}{0.84 $\pm$ 0.12} &
\multicolumn{3}{c}{0.73 $\pm$ 0.11} & \multicolumn{3}{c}{0.80 $\pm$ 0.16} & \multicolumn{3}{c}{0.64 $\pm$ 0.10} &
\multicolumn{3}{c}{0.55 $\pm$ 0.10} \\

c(\hbeta) & \multicolumn{3}{c}{0.62 $\pm$ 0.24} &
\multicolumn{3}{c}{0.87 $\pm$ 0.21} & \multicolumn{3}{c}{$-$} & \multicolumn{3}{c}{0.98 $\pm$ 0.13} &
\multicolumn{3}{c}{1.10 $\pm$ 0.12} \\

E$_{\rm B-V}$ & \multicolumn{3}{c}{0.48 $\pm$ 0.19} &
\multicolumn{3}{c}{0.67 $\pm$ 0.17} & \multicolumn{3}{c}{$-$} & \multicolumn{3}{c}{0.75 $\pm$ 0.10} & \multicolumn{3}{c}{0.85 $\pm$ 0.09} \\
\hline
\end{tabular}
\end{minipage}
\end{table*}

\begin{table*}
\label{table3b}
\begin{minipage}{16cm}
\contcaption{}
\begin{tabular}{lccccccccccccccc}
\hline
\noalign{\smallskip}
 & \multicolumn{3}{c}{Slit 5c} & \multicolumn{3}{c}{Slit 5d} 
& \multicolumn{3}{c}{Slit 6a} & \multicolumn{3}{c}{Slit 6b} &
\multicolumn{3}{c}{Slit 6c} \\ 
Line (\AA) & F\footnote{Observed fluxes normalized to F(H$\alpha$)=100 and
uncorrected for interstellar extinction.} & I\footnote{Observed fluxes normalized to F(H$\alpha$)=100 and
corrected for interstellar extinction.}  & S/N\footnote{Numbers represent the signal to noise ratio of the quoted fluxes.} & F & I & S/N
& F & I & S/N & F & I & S/N & F & I & S/N \\
\hline
\hbeta\ 4861 & 17 & 35 & 6 & 19 & 35 & 6 & 16 & 35 & 5 & $-$ & $-$ & $-$ & 18 & 35 & 5 \\
\oxygen\ 4959 & $-$ & $-$ & $-$ & $-$ & $-$ & $-$ & 7 & 15 & 3 & 11 & $-$ & 3 & $-$ & $-$ & $-$ \\
\oxygen\ 5007 & $-$ & $-$ & $-$ & 7 & 12 & 3 & 18 & 36 & 4 & 37 & $-$ & 6 & 14 & 25 & 5 \\
\nitrogen\ 5755 & $-$ & $-$ & $-$ & $-$ & $-$ & $-$ & $-$ & $-$ & $-$ & $-$ & $-$ & $-$ & $-$ & $-$ & $-$ \\
\oi\ 6300 & 15 & 17 & 7 & 7 & 7 & 7 & 21 & 23 & 7 & 25 & $-$ & 7 & 14 & 15 & 7 \\
\oi\ 6364 & 5 & 5 & 3 & 2 & 2 & 3 & 7 & 8 & 3 & 8 &  $-$  & 2 & 5 & 5 & 3 \\
\nitrogen\ 6548 & 15 & 16 & 7 & 12 & 12 & 5 & 15 & 15 & 6 & 28 & $-$ & 7 & 11 & 11 & 5 \\
\ha\ 6563 & 100 & 100 & 39 & 100 & 100 & 35 & 100 & 100 & 28 & 100 & $-$ & 27 & 100 & 100 & 33 \\
\nitrogen\ 6584 & 50 & 50 & 20 & 41 & 41 & 14 & 45 & 45 & 16 & 65 & $-$ & 19 & 38 & 38 & 14 \\
\sulfur\ 6716 & 48 & 48 & 21 & 37 & 37 & 13 & 54 & 53 & 17 & 58 & $-$ & 17 & 38 & 38 & 16 \\
\sulfur\ 6731 & 33 & 33 & 15 & 25 & 25 & 10 & 37 & 37 & 15 & 36 & $-$ & 13 & 31 & 31 & 14 \\
\hline
Absolute \ha\ flux\footnote{In units of \fluxa.} & \multicolumn{3}{c}{176.8} &
\multicolumn{3}{c}{91.1} & \multicolumn{3}{c}{32.4} &
\multicolumn{3}{c}{26.5} & \multicolumn{3}{c}{25.8} \\

\sulfur/\ha\ & \multicolumn{3}{c}{0.81 $\pm$ 0.07} &
\multicolumn{3}{c}{0.62 $\pm$ 0.08} & \multicolumn{3}{c}{0.88 $\pm$ 0.08} & \multicolumn{3}{c}{0.94 $\pm$ 0.10} & \multicolumn{3}{c}{0.68 $\pm$ 0.07} \\
F(6716)/F(6731) & \multicolumn{3}{c}{1.44 $\pm$ 0.10} &
\multicolumn{3}{c}{1.47 $\pm$ 0.15} & \multicolumn{3}{c}{1.46 $\pm$ 0.10} &
\multicolumn{3}{c}{1.58 $\pm$ 0.12} & \multicolumn{3}{c}{1.25 $\pm$ 0.09} \\

\nitrogen/\ha\ & \multicolumn{3}{c}{0.65 $\pm$ 0.10} &
\multicolumn{3}{c}{0.53 $\pm$ 0.11} & \multicolumn{3}{c}{0.60 $\pm$ 0.11} & \multicolumn{3}{c}{0.93 $\pm$ 0.14} & \multicolumn{3}{c}{0.49 $\pm$ 0.10} \\

c(\hbeta)\footnote{The logarithmic extinction is derived by c =
1/0.348$\times$log((\ha/\hbeta)$_{\rm obs}$/2.85).} & \multicolumn{3}{c}{0.89 $\pm$ 0.16} &
\multicolumn{3}{c}{0.78 $\pm$ 0.13} & \multicolumn{3}{c}{0.98 $\pm$ 0.23} & \multicolumn{3}{c}{$-$} & \multicolumn{3}{c}{0.83 $\pm$ 0.13} \\
E$_{\rm B-V}$\footnote{The interstellar reddening was measured from the relation E$_{\rm B-V}$ $\approx$ 0.77c (\citealt{Osterbrock2006})} & \multicolumn{3}{c}{0.68 $\pm$ 0.12} &
\multicolumn{3}{c}{0.60 $\pm$ 0.10} & \multicolumn{3}{c}{0.75 $\pm$ 0.18} & \multicolumn{3}{c}{$-$} & \multicolumn{3}{c}{0.64 $\pm$ 0.09} \\
\hline

 & \multicolumn{3}{c}{Slit 6d} &
\multicolumn{3}{c}{Slit 7a} & \multicolumn{3}{c}{Slit 7b} &
\multicolumn{3}{c}{Slit 8a} & \multicolumn{3}{c}{Slit 8b} \\ 
Line (\AA) & F & I & S/N & F & I & S/N & F & I & S/N & F & I & S/N & F
& I & S/N \\
\hline 
\hbeta\ 4861 & 15 & 35 & 4 & 15 & 35 & 4 & $-$ & $-$ & $-$ & $-$ & $-$ & $-$ & $-$ & $-$ & $-$ \\
\oxygen\ 5007 & 33 & 73 & 8 & $-$ & $-$ & $-$ & $-$ & $-$ & $-$ & $-$ & $-$ & $-$ & $-$ & $-$ & $-$ \\
\oi\ 6300 & 13 & 15 & 3 & 11 & 12 & 5 & 14 & $-$ & 5 & 21 & $-$ & 4 & 6 & $-$ & 4 \\
\oi\ 6364 & $-$ & $-$ & $-$ & $-$ & $-$ & $-$ & $-$ & $-$ & $-$ & 12 & $-$ & 2 & $-$ & $-$ & $-$ \\
\nitrogen\ 6548 & 16 & 17 & 6 & 19 & 20 & 8 & 41 & $-$ & 5 & $-$ & $-$ & $-$ & 19 & $-$ & 3 \\
\ha\ 6563  & 100 & 100 & 27 & 100 & 100 & 35 & 100 & $-$ & 19 & 100 & $-$ & 19 & 100 & $-$ & 15 \\
\nitrogen\ 6584 & 48 & 48 & 16 & 58 & 57 & 22 & 60 & $-$ & 14 & 32 & $-$ & 4 & 45 & $-$ & 7 \\
\sulfur\ 6716 & 37 & 37 & 13 & 41 & 41 & 18 & 54 & $-$ & 13 & 44 & $-$ & 12 & 32 & $-$ & 6 \\
\sulfur\ 6731 & 31 & 30 & 11 & 26 & 26 & 15 & 32 & $-$ & 10 & 32 & $-$ & 9 & 24 & $-$ & 4 \\
\hline 
Absolute \ha\ flux & \multicolumn{3}{c}{26.1} &
\multicolumn{3}{c}{85.9} & \multicolumn{3}{c}{4.1} &
\multicolumn{3}{c}{24.3} & \multicolumn{3}{c}{8.8} \\

\sulfur/\ha\ & \multicolumn{3}{c}{0.67 $\pm$ 0.08} &
\multicolumn{3}{c}{0.67 $\pm$ 0.07} & \multicolumn{3}{c}{0.86 $\pm$ 0.12} & \multicolumn{3}{c}{0.76 $\pm$ 0.10} & \multicolumn{3}{c}{0.56 $\pm$ 0.16} \\

F(6716)/F(6731) & \multicolumn{3}{c}{1.22 $\pm$ 0.11} &
\multicolumn{3}{c}{1.58 $\pm$ 0.11} & \multicolumn{3}{c}{1.68 $\pm$ 0.17} & \multicolumn{3}{c}{1.35 $\pm$ 0.15} & \multicolumn{3}{c}{1.35 $\pm$ 0.31} \\

\nitrogen/\ha\ &\multicolumn{3}{c}{0.65 $\pm$ 0.12} &
\multicolumn{3}{c}{0.78 $\pm$ 0.11} & \multicolumn{3}{c}{1.01 $\pm$ 0.22} & \multicolumn{3}{c}{0.32 $\pm$ 0.08} &
\multicolumn{3}{c}{0.64 $\pm$ 0.22} \\

c(\hbeta) & \multicolumn{3}{c}{1.10 $\pm$ 0.22} &
\multicolumn{3}{c}{1.10 $\pm$ 0.25}& \multicolumn{3}{c}{$-$}& \multicolumn{3}{c}{$-$} &
\multicolumn{3}{c}{$-$} \\

E$_{\rm B-V}$ & \multicolumn{3}{c}{0.85 $\pm$ 0.17}
&\multicolumn{3}{c}{0.85 $\pm$ 0.19} & \multicolumn{3}{c}{$-$} & \multicolumn{3}{c}{$-$} & \multicolumn{3}{c}{$-$} \\
\hline
\end{tabular}
\end{minipage}
\end{table*}

\begin{figure*}
\label{2D_diag}
\centering
\includegraphics[scale=0.7]{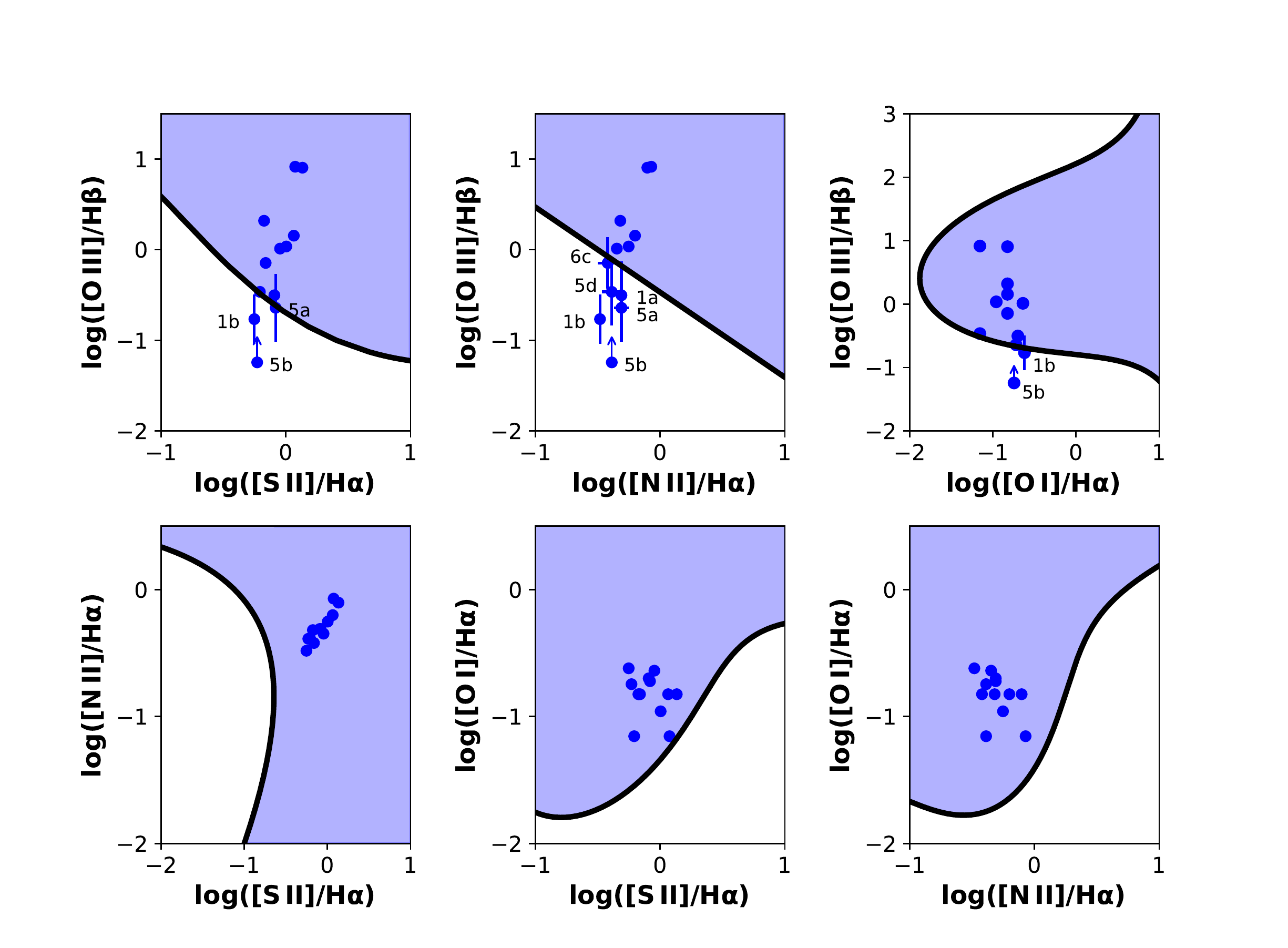}
\caption{2D diagnostic curves of \citet{Kopsacheili2020} together with the results extracted by our spectroscopic data (blue points). The blue region is the locus that indicates shock excited gas emission.}
\end{figure*}

\subsection{Comparison with other wavelengths}\label{sec:comparison}

G132.7+01.3 has been studied in $\gamma$-rays, X-rays, radio, infrared and mm wavelengths \citep[e.g.][ see also Introduction]{Routledge1991,Fesen1995,Lazendic2006,Zhou2016, Rho2021}. Here we present the main results and conclusions regarding the properties and evolution of HB3 as extracted by the comparison of its emission properties in different electromagnetic bands. In Fig. \ref{fig9} the contours of the emission that the remnant reveals in the radio, X-rays, CO lines and $\gamma$-rays are superposed on the \hnii~image of HB3 from our observations.

The optical morphology of the remnant displays a strong correlation with its radio emission at both large and small spatial scales, signifying that they have the same origin  (Fig. \ref{fig9}a). In particular, the outer bright \hnii~filaments at the western portion of the remnant follow both the geometry and brightness of the radio emission that declines gradually, moving from the west to the northwest region of HB3. This advocates that the remnant encounters a denser ambient medium in the west part than in the northwest. Intriguingly, both in radio and optical the remnant reveals two antisymmetric local protrusions in west and east giving the morphological impression of two lobes in the overall shape of the remnant.  In the southern portion of HB3 even though the optical and radio display a similar morphology, the two  bands are not in good correlation . In particular, the optical emission in the south of HB3 displays a relatively faint and diffuse emission, while the radio emission in this portion is equally bright as compared to the west region of HB3. This could be attributed either to dust absorption affecting the optical photons and/or to a high local blast wave velocity that enhances the non-thermal synchrotron emission at this region.  Finally,  the eastern region of HB3 the radio emission of the remnant roughly follows the bright filament in \hnii~marking the border between the SNR and the ambient H{\sc ii}/MC complex. 

Regarding the inner, thermal X-ray emission of the remnant (Fig. \ref{fig9}b) -a property that classifies HB3 as a MMSNR \citep{Rho1998,Lazendic2006}- it is lying close to the center of the optical image of HB3 being centralized at about $ \left( \alpha, \delta \right)= \left(2^h~18^m, 62^o~40'  \right)$. The remnant's X-ray brightness declines moving from the center to the outer portions of HB3, following initially a roughly spherical distribution but at the same time it displays two large anti-asymmetric wings towards the northeast and south that cover the whole optical image of the remnant in these regions. The overall X-ray morphology of HB3 is shifted towards the eastern region of the remnant where it intersects with the  {H\sc ii}/MC complex, something that advocates that the mixed morphology properties of HB3 are related to the interaction of the remnant with the surrounding cloud. Such a hypothesis is aligned to evaporating clouds model \citep{WhiteLong91} and the reflected shock model \citep{Chen2008} for the formation of MMSNRs as both demand the interaction of the SNR with a surrounding dense cloud (see also Sect. \ref{sec:MMSRNorigin}).

The molecular gas in the vicinity of HB3 as detected by millimeter observations of CO lines  \citep{Zhou2016}, coincides with the bright \hnii\ and \sii~ southeast region of the SNR, lying about $ \left( \alpha, \delta \right)= \left(2^h~18^m~30^s, 62^o~20'  \right)$. Given that the optical emissivity of the shocked gas goes as $\rho^2$ where $\rho$ its mass density, it is inferred that in this region the SNR blast wave is propagating in a dense environment. This result offers an additional evidence towards the conclusion that HB3 is currently interacting with the W3 H{\sc ii} region/MC complex \citep[see also][for an additional argumentation]{Zhou2016, Routledge1991}. 

Finally, the $\gamma$-ray emission discovered in the vicinity of HB3 \citep[][; see also Fig. \ref{fig9}d]{Tsygankov2016, Katagiri2016} is spatially correlated with the bright $^{12}CO$ (J = 1-0) emission and it is adjacent to the  southeast region of the remnant. Such a result favors for a hadronic origin of the  $\gamma$-ray photons resulted by interactions between particles accelerated in the SNR and the molecular cloud. 

All above evidence advocate that HB3 is indeed expanding in the vicinity of the W3  H{\sc ii} region/MC complex and currently is interacting with it at its eastern portion.  This interaction seems to substantially affect the morphology of the remnant as well as its emission properties all over the electromagnetic spectrum. The borders between the SNR and the surrounding cloud are well justified by the radio observations and the bright \hnii\ filaments at the eastern region of the remnant.

\section{Discussion}

\subsection{Properties of the optical filaments}

Having an angular diameter of $\sim$90\arcmin$\times$120\arcmin, HB3 is among the largest Galactic SNRs known. Furthermore, assuming a distance of $\sim$2 kpc, this means a linear diameter of 52 $\times$72 pc, indicative of an evolved SNR being in the later Sedov--Taylor   or in the   radiative phase.
The composite \hnii\ image provides the best description of the morphology of G132.7+01.3 in its full extent making possible the identification of all of its detailed structures. Emission from the \sulfur~$\lambda\lambda$~6716,6731 doublet lines and the \oiii~$\lambda$5007 line has also been detected matching in shape and position the \hnii~emission, though being more diffuse and less filamentary. Furthermore, the prominent \oiii\: emission indicates the existence of fast shocks travelling through the ionized gas all around the remnant. (\citealt{cox1985}). 

The \sulfur/H$\alpha$ emission line ratio measured from the optical images of the Area A (Table \ref{table2}) does not exceed the threshold (0.4) of supernova remnants. We thus argue that this particular area does not belong to G132.7+01.3, but it is associated with the H{\sc ii} region. Note that the area A is also the brightest in the H$\alpha$ emission line (see Table~\ref{table2}). All other areas display \sulfur /H$\alpha>$ 0.4, except area~S7 for which a ratio of 0.35 is determined. Note that the width of the \ha+\nii~filter used for the computation of \ha~emission is large enough to transmit the \nitrogen~6584~ \AA\ line. This extra emission is likely responsible for the lower \sulfur /H$\alpha$ ratio measured for the area~S7.   

Regarding our spectroscopic observations of eight filaments distributed over the entire remnant, the \sii/H$\alpha$ ratios are clearly higher than 0.4. The distribution of our spectra in the H$\alpha$/\sii~--H$\alpha$/[N{\sc ii}] space is consistent with the locus of shock-heated gas in SNRs and well separated from the UV-dominated PNe and H{\sc ii} regions \citep[e.g.][]{Sabbadin1977, Meaburn2010, Leonidaki2013, Sabin2013, Akras2020}. The strong \oi~$\lambda$6300 \AA ~emission in most of the filaments provides an extra confirmation of their shock-heated origin. 

In Fig.~\ref{2D_diag}, we present the newly developed diagnostic diagrams between shock-heated and UV~photo-ionized regions for different combinations of emission line ratios \citep{Kopsacheili2020}. Our observed \sii~/H$\alpha$, [N{\sc ii}]/H$\alpha$ and [O~{\sc i}]/H$\alpha$ ratios from all eight filaments and different apertures satisfy the criteria and lie within the regime of expected shock-heated regions. 

However, a small discrepancy is observed when the [O~{\sc iii}]/H$\beta$ criterion is used for slits 1 and 5. The measured intensities of the emission line(s) of these slits are highly uncertain, with signal-to-noise values of 2-4 (see Table~\ref{table3}). This is a possible interpretation why regions corresponding to Slit 1 and 5 fall outside the expected locus of shock-heated regions in the [O~{\sc iii}]/H$\beta$ diagnostic diagrams. This can also be seen from the [\ion{O}{III}]~5007\AA\ emission which is very weak or even absent at the position of the slits 1 and 5 (see RGB Figure \ref{fig3} of HB3). A possible contribution in the \hb~emission from the nearby {H\sc ii}~region W3 in slit 5, that would result in lower [\ion{O}{III}]/\hb~ratio, cannot be ruled out. We point out however that the new diagnostics do not account for shocks with velocities lower than 100~\vel~(only models with shock velocities between 100 and 1000~\vel~are considered, \citealt{Kopsacheili2020}). Furthermore, we note that all the diagnostics are characterized by incompleteness, which is higher in the case of [N~{\sc ii}]/H$\alpha$--[O~{\sc iii}]/H$\beta$ \citep[for more details see, ][]{Kopsacheili2020}.

\subsection{Origin and evolution of HB3}
\label{sec:HB3_origin}

Adopted from the optical image extent of HB3 ($ \sim 90' \times 120'$) and a distance of $d=$ 2 kpc, the average radius of the remnant is $R_s=$ 31 pc. Assuming further that the remnant is still in the Sedov phase and the magnetic field pressure is negligible, the values of the preshocked ambient density ($n_o$)  and the current shock’s velocity ($V_s$)   can be constrained by using the relation  \citep{Dopita1979}:  
\begin{equation}\label{eq:nsII}
{\rm n_{[S {\sc II}]} \simeq\ 45\ \left( \frac{n_o}{\rm cm^{-3}} \right) \left( \frac {V_{\rm s}}{100~km~s^{-1}} \right)^2}~{\rm cm^{-3}},
\end{equation}
\\
where $n_{\rm [S II]}= 160 \pm 80~\rm cm^{-3}$,  extracted from the \sii~ diagnostic lines.  From Eq. (\ref{eq:nsII}), we get: $\left( \frac{n_o}{\rm cm^{-3}} \right) \left( \frac {V_{\rm s}}{100~km~s^{-1}} \right)^2= 3.6 \pm 1.8$ . Such a result is in agreement with the independent estimations of HB3 shock velocity and pre-shock density by \citet{Lazendic2006} who found $\left( \frac{n_o}{\rm cm^{-3}} \right) \left( \frac {V_{\rm s}}{100~km~s^{-1}} \right)^2= \left( \frac{0.32}{\rm cm^{-3}} \right) \left( \frac {340}{100~km~s^{-1}} \right)^2= 3.7 \pm 1.4$. 
Additional constraints on the ambient medium density -and thus on the current shock velocity- we get by estimating the HB3 column density. Using the relation of \cite{Ryter1975}, we obtain an N$_{\rm H}$~of $3.3 (\pm1.5) \times 10^{21}~{\rm cm}^{-2}$~and $6.0 (\pm1.7) \times 10^{21}~{\rm cm}^{-2}$~for the minimum and maximum c(H$\beta$) values  as calculated from our spectra, respectively. The extracted column density range is consistent to the estimated values of N$_{\rm H} \sim 6.2 \times 10^{21}$~cm$^{-2}$ \citep{Dickey1990} and N$_{\rm H} \sim 7.7 \times 10^{21}$~cm$^{-2}$  \citep{Kalberla2005} in the direction of HB3, and the one extracted by  \cite{Fesen1995} using their optical spectra in HB3 western rim, who found  N$_{\rm H} \sim 5.7\times 10^{21}~{\rm cm}^{-2}$. The corresponding pre-shocked ambient medium number density range is $0.5 - 1.0 ~d_2^{-1} \rm cm^{-3}$ having a mean value of $n_o= \left( 0.82 \pm 0.15 \right)~d_2^{-1}$ $\rm cm^{-3}$, where $d_2$ the distance of HB3 in units of 2~kpc.   Using Eq. (\ref{eq:nsII}) we  extract a mean shock wave velocity of $V_s = \left( 208 \pm 54 \right)~d_2^{0.5}\rm km~s^{-1}$. Employing the aforementioned results extracted by our optical observations on the  Sedov solution, we find an explosion energy for HB3 equal  to $E~\sim~2.1~\times~10^{46}~\left( \frac{R_s}{pc} \right)^3~\left( \frac{n_o}{cm^{-3}} \right)~\left( \frac {V_{\rm s}}{100~km~s^{-1}} \right)^2 = \left(2.3 \pm 1.1\right)~\times~10^{51} d_2^3~\rm erg$ \citep{Ostriker1988} while the SNR age is estimated to be $t \sim \left( 5.8 \pm 1.5 \right) \times 10^4~d_2^{0.5}$ yrs.

The large size of the remnant and its bright optical line emission indicate that HB3 has evolved beyond the SNR adiabatic phases and currently is in the Pressure Driven Snowplow (PDS) phase. In order to assess the correctness of this  assumption, we estimate the transition radii from the Sedov to the PDS phase extracted by \cite{Cioffi1988}:
\begin{equation}\label{eq:RPDS}
{\rm R_{PDS}= 14.0 \left( \frac{E}{10^{51}~erg} \right)^{\frac{2}{7}} \left( \frac{n_o}{cm^{-3}} \right)^{-\frac{3}{7}} \zeta_m^{-\frac{1}{7}}},
\end{equation}
\\
where $E$ is the SN energy and $\zeta_m$ a constant equal to unity for solar metalicity. In the PDS phase the SN energy is found to be $E= 6.8 \times 10^{43} \left(\frac{n_o}{cm^{-3}}\right)^{1.16} \left( \frac{V_s}{km~s^{-1}} \right)^{1.35}  \left( \frac{R_s}{pc} \right)^{3.16} \zeta_m^{0.161} = \left( 3.7 \pm 1.5 \right) \times 10^{51} d_2^{2.7}$~erg \citep{Cioffi1988} for which the corresponding transition radius is $R_{\rm PDS} = \left( 22.2 \pm 3.1 \right) d_2^{1.2}$~pc. The transition radius is slightly smaller that the current radius of HB3 ($Rs \sim 31 d_2$~pc) something that indicates that the remnant has evolved into the PDS phase or is currently in the transition phase between the Sedov and PDS stages. Intriguingly, this finding is commonly met in the class of MMSNRs \citep{Shelton1999}. The corresponding time in which HB3 started to enter into the PDS phase is $t_{\rm PDS} = 1.33  \times 10^4 \left( \frac{E}{ 10^{51}~erg} \right)^{\frac{3}{14}}\left( \frac{n_0}{cm^{-3}} \right)^{-\frac{4}{7}} \zeta_m^{-\frac{5}{14}} = \left( 2.0 \pm 0.3 \right)\times10^4~d_2^{1.15}$~yrs, while the current age of the remnant was found to be $t=~\frac{3}{4} t_{\rm PDS} \left[ \left( \frac{R_s}{R_{\rm PDS}} \right)^{\frac{10}{3}} + \frac{1}{3} \right]=  \left( 5.1~\pm 2.1 \right)~\times~10^4$~yrs. These timescales indicate that HB3 has spent at least the 30\% of its lifetime in the PDS phase. 

\subsection{On the mixed morphology of HB3}
\label{sec:MMSRNorigin}

The previous approach provides a first order approximation on the current evolutionary properties of the remnant as well as on its parent SN explosion. Nevertheless, it neglects the following two major facts: a) the ambient medium properties around HB3 deviate substantially from the homogeneity revealing large spatial gradients. In particular, our optical study confirmed that HB3 interacts in its eastern region with a dense H{\sc ii}/MC complex something that affects its morphology, kinematics and emission properties. Thus, the evolution of the remnant cannot be accurately approached by one-dimensional, spherically symmetric models; b) HB3 has been classified as a mixed morphology SNR being characterized by the coexistence of an optical/radio bright shell with a  centrally peaked thermal X-ray emission. These SNR properties cannot be explained by the canonical SNR evolution models and therefore,  additional `ingredients’ on its evolutionary history are required.

\citet{Lazendic2006} provided an explanation on the mixed morphology properties of HB3 by invoking the so called `evaporating clouds model' \citep[e.g.][]{WhiteLong91,Slavin2017, Zhang2019} and adopting  a $C \over \tau$ ratio of 3-5, where $C$ is the mass ratio between the clouds and the intercloud medium and $\tau$ is the ratio of the cloud evaporation timescale to the current SNR age. 

Here we discuss an alternative evolutionary scenario of HB3 that potentially could led on its mixed morphology properties based on the `reflected shock model'.  According to this model  MMSNRs are formed by the action of a reflected shock, triggered by the collision of the SNR with the density walls of a  pre-existing circumstellar cavity sculptured by the mass outflows of the progenitor star \citep{Dwarkadas2005, Chen2008, Zhang2015}.

If this was the case of HB3, the size of the formed wind-blown cavity should be comparable to the current size of the remnant. This fact provides us an estimation of the progenitor mass  by employing the linear relation of  \citet{Chen2013}:
\begin{equation}\label{eq:Rd_Mms}
{\rm R_{b} \simeq\ \left( 1.22 \pm 0.05 \right) \times \left( \frac{M_{ms}}{M_{\odot}} \right)  - \left(9.16 \pm 1.77 \right)~pc }.
\end{equation}
where $R_b$ the radius of the wind blown cavity and $M_{ms}$ the stellar main sequence mass. Considering that the current radius of HB3 is $\sim 31~d_2$ pc  we estimate that the remnant resulted by a progenitor star with a main sequence mass of about $M_{ms}= 33.9 \pm 2.0 ~\rm M_{\odot}$. According to the stellar evolution theory this star most likely passed through the Wolf-Rayet phase before its final explosion. If true, then HB3 resulted by a Type Ib/c SN \citep[][and references therein]{Smith2014}. 

According to this scenario, the remnant initially evolved into the wind blown cavity of a size $R_b \sim 31$~pc, shaped by the progenitor star. Describing the SN ejecta  density profile with a power law of $\rho_{ej} \propto r^{-n}$, with n=7 \citep[appropriate for Type I SNe;][]{Chevalier1981}, and assuming energy conservation, the evolution of the SNR within the wind cavity till its collision with the density walls of the circumstellar structure  can be approached by  employing the self-similar solution of \citet{Chevalier1982}. The radius ($R_s$) and velocity ($V_s$) of a SNR expanding in an wind blown cavity with a density profile $\rho_{AM}=q~r^{-s}$, where s=2  and $q= \frac {\bm \dot{M_w}} { 4 \pi u_w }$ and  are given by:

\begin{equation}\label{eq:Rs_Chev}
{\rm R_{s} = 1.3 \times \left( \frac{A~g^n}{q} \right)^{\frac{1}{n-s}}  t^{\frac{n-3}{n-s}}        }.
\end{equation}

\begin{equation}\label{eq:Rs_Chev}
{\rm V_{s} = 1.3 \times  \left(\frac{n-3}{n-s} \right) \times   \left( \frac{A~g^n}{q} \right)^{\frac{1}{n-s}}  t^{\frac{s-3}{n-s}}        }.
\end{equation}
where for n= 7 we have A= 0.27 and $g= \left[ \left( \frac{25}{21 \pi}\right) \left(\frac{E_{ej}^2}{M_{ej}}\right) \right]^\frac{1}{7}$  and t is the age of the remnant.  By assuming the canonical energy for the SN explosion of $E_{ej}= 10^{51}$~erg, an ejecta mass of $M_{ej}= 9~\rm~M_{\odot}$ \citep[as expected to be the final mass of a  $34~\rm~M_{\odot}$ progenitor e.g.][]{Meyer2021} and adopting typical WR wind properties $\left({\bm \dot{M}} \over  10^{-5}~\rm~M_{\odot}~yr^{-1} \right) \times  \left(\dot{u_w} \over  10^{3}~\rm~km~s^{-1} \right)^{-1}= 1$  we find that the remnant reached with the density walls of the wind blown cavity after $t\approx 4500$~yrs  of evolution having an incident shock velocity of $Vs \approx 5400~\rm~km~s^{-1}$. 

At the moment of the collision between the SNR and the density wall a pair of shocks is formed: the transmitted shock that starts to penetrate the density wall and the reflected shock that moves inwards and shocks the SN ejecta. Due to the high wall-to-cavity density contrast the transmitted shock is expected to decelerate substantially \citep{Dwarkadas2005,Dwarkadas2007, Chiotellis2012, Chiotellis2013}. In particular, the transmitted shock velocity ($v_t$) right after the collision of the SNR  with the cavity walls as a function of the incident shock velocity ($Vs$) is given by \citep{Sgro1975}:
\begin{equation}\label{eq:Vt}
{\rm V_{t} = \left( \frac{\beta}{A} \right)^{\frac{1}{2}}\times V_s}.
\end{equation}
where $A$ is the wall to cavity density contrast and  $\beta$ the pressure ratio of the post-transited-shock gas to the the post-incident-shock-gas.  The ratio $\beta$ is related to the density contrast between the post-reflected-shock gas and post-incident shock gas ($A_r$) as:
\begin{equation}\label{eq:beta}
{\rm \beta =  \frac{4 A_r -1}{4- A_r} },
\end{equation}
and $A_r$ in turn is related to $A$ as:
\begin{equation}\label{eq:A}
{\rm A =  \frac{3 A_r \left( 4 A_r -1 \right)}{\left[ \left[3A_r \left(4-A_r \right) \right]^{\frac{1}{2}}-\sqrt{5} \left( A_r -1 \right) \right]^2} }.
\end{equation}

For the adopted wind parameters the number density at the outer region of the wind cavity will be in the order of $n_{cav.} \sim~10^{-3}~\rm cm^{-3}$. The cavity wall density cannot be described by a single value as it is strongly related to the ambient medium density, that as discussed above deviates substantially from homogeneity. Adopting an ambient medium of $n_o= 0.5 -1 ~\rm cm^{-3}$, as extracted by the column density estimations, for the western portion of the remnant, and a $n_{cloud} \sim 10^1 -10^2 ~\rm cm^{-3}$ for the eastern cloud and considering that the ambient medium has been compressed by at least a factor of 4, we  get $A \sim 10^3 - 10^4$ and $A \sim 10^4 - 10^5$ for the western and eastern regions of the remnant, respectively. Importing these values to Eq. (\ref{eq:Rs_Chev}) - (\ref{eq:A}) we get a transmitted shock velocity at west of $V_t \approx 130 - 400~\rm km~s^{-1}$ \citep[consistent to the estimations of][]{Lazendic2006} while to the eastern region a much slower one in the order of $V_t \approx 40 - 130~\rm km~s^{-1}$. Regarding the reflected shock, its initial velocity is given:

\begin{equation}\label{eq:Vr}
{\rm V_{r} = \frac{1}{4} \left[3-\left(\frac{15 A_r}{4-A_r} \right)^\frac{1}{2} \right] V_s},
\end{equation}
from which we get $V_r \approx 2300 - 2500~\rm km~s^{-1}$ at the east and $V_r \approx 2500 - 2600~\rm km~s^{-1}$ at the west of HB3. A shock wave with such a high velocity will compress and heat the gas to temperatures that becomes X-ray bright. 

Such a scenario seems  consistent with the observed properties of HB3, since in principle can explain the low expansion velocities of HB3 forward shock and at the shame time its central X-ray emission.  As the remnant evolved further the transmitted shock penetrated more the density wall and it got further decelerated up to its current expansion velocities. Simultaneously, the reflected shock reached the center of the remnant producing the overall centrally X-ray peak morphology of HB3. In order to verify  whether such a scenario could indeed reproduce the current properties of HB3 detailed hydrodynamic modeling is required. 

\section{Conclusions}

\begin{enumerate}
  \item HB3 (G 132.7$+$1.3), which is among the largest supernova remnants, was studied in optical wavelengths in its full extent for the first time. Filamentary and diffuse structures are both present in its optical images. The \sulfur\ emission of HB3, though fainter, displays a similar morphology to that of the \hnii\ emission-line image, tracing all the filamentary structures. On the other hand, the \oiii\ emission line, displays a different filamentary morphology compared to the former line images, indicating different shock velocities and physical conditions.
  
  \item Spectroscopic data from eight filaments distributed across the whole remnant exhibit strong \ha, \nii~$\lambda\lambda$6548,6584, \sii~$\lambda\lambda$6716,6731 and \oi~$\lambda\lambda$6300,6363 emission lines and low-to-moderate emission from the \oiii~$\lambda\lambda$4959,5007 lines. These findings indicate shock waves with velocities $<$100~km~s$^{-1}$ for most of the filaments with an exception of slit~2, which has higher \oiii/\hb~ratio and shock velocity up to 120~km~s$^{-1}$. The \sii~lines ratio implies an electron density $<$240~cm$^{-3}$.
  
  \item The observed emission line ratios are also found to be in agreement with the diagnostic diagrams that distinguish shock-heated gas (SNRs) and UV-photoionized gas ({H\sc ii} regions) verifying the origin of the optical emission from shock-heated gas.

  \item We compared the \ha+\nii~image extracted by our observations with the emission that HB3 displays in the radio, X-rays, millimeter and $\gamma$-rays bands. Overall, the radio emission of HB3 is co-spatial with the bright optical filaments of the remnant something that indicates the same physical origin of the two radiations and marks the border of the SNR in respect to the adjacent cloud. The X-ray emission peaks roughly close to the optical center of the remnant but its overall distribution tends to be shifted towards the east, advocating that there is a link between the interaction of the remnant with the {H\sc ii}/MC complex and the mixed morphology properties of HB3. Finally, regarding the CO emission line and  $\gamma$-ray emission there are co-spatial with the bright southeast optical filament of the remnant something that offers an additional evidence on the interaction of HB3 with the W3 complex. 
  
 \item Based on the  \sii~ diagnostic lines and column densities obtained by our observations, we estimated the current shock velocity and preshock ambient medium density of HB3 and we found to be $V_s = 208 \pm 54~\rm km~s^{-1}$ and $n_o = 0.82 \pm 0.15~\rm cm^{-3}$, respectively. Regarding the evolutionary state of HB3, we show that the remnant has most likely passed into the PDS phase or it is in transition between the Sedov and PDS stages. Under this assumption, the SN energy of HB3 was found to be $\left( 3.7 \pm 1.5 \right) \times 10^{51}$~erg and its current age $\left(5.2 \pm 2.1\right) \times 10^4$~yrs.   
 \item We discussed the mixed morphology properties of HB3, under the reflected shock model. We show that the overall properties of the remnant, namely its optical/radio bright shocks and its central X-ray emission, can in principle be explained considering that the remnant is currently interacting with the density walls of a wind blown cavity shaped by a progenitor  star with initial mass of $\sim 32 - 36~\rm M_{\odot}$.
\end{enumerate}

\section*{Acknowledgements}

This research is co-financed by Greece and the European Union (European Social Fund-ESF) through the Operational Programme “Human Resources Development, Education and Lifelong Learning 2014-2020” in the context of the project “On the interaction of Type Ia Supernovae with Planetary Nebulae” (MIS 5049922). A.C. acknowledge the support of this work by the project ``PROTEAS II'' (MIS 5002515),which is implemented under the Action ``Reinforcement of the Research and Innovation Infrastructure'', funded by the Operational Programme ''Competitiveness, Entrepreneur- ship and Innovation'' (NSRF 2014–2020) and co-financed by Greece and the European Union (European Regional
Development Fund). We thank (a) Pat Slane for providing us with the ROSAT PSPC X-ray image in fits format, (b) Xin Zhou for providing us with the CO images in fits format and (c) Hideaki Katagiri for providing us with the $\gamma$-ray image in fits format.
This work is based on observations made with the ``Aristarchos'' telescope operated on the Helmos Observatory by the Institute of Astronomy, Astrophysics, Space Applications and Remote Sensing of the National Observatory of Athens. Skinakas Observatory is a collaborative project of the University of
Crete, the Foundation for Research and Technology-Hellas and the
Max-Planck-Institut f\"ur Extraterrestrische Physik.  Observatory is a collaborative project of the University of
Crete, the Foundation for Research and Technology-Hellas and the
Max-Planck-Institut f\"ur Extraterrestrische Physik. This research
made use of data from SuperCOSMOS \ha\ Survey (AAO/UKST), from the ATNF Pulsar Catalogue and from the NASA/IPAC Infrared Science Archive.

\section*{Data Availability}

The data underlying this article will be shared on reasonable request to the corresponding author.


\bibliographystyle{mnras}
\bibliography{HB3_bibl}

\bsp	
\label{lastpage}
\end{document}